\documentclass[manuscript]{acmart}

\AtBeginDocument{%
  }

\setcopyright{acmlicensed}
\copyrightyear{2025}
\acmYear{2025}
\acmDOI{XXXXXXX.XXXXXXX}
\acmConference[ACM TSAS]{ACM Transactions on Spatial Algorithms and Systems}{June 03--05,
  2025}{Woodstock, NY}
\acmISBN{978-1-4503-XXXX-X/2025/06}

\newcommand\vldbavailabilityurl{https://github.com/MajidSas/GS-QA}

\usepackage{lipsum}
\usepackage{multirow}
\usepackage{graphicx}
\usepackage{threeparttable}
\usepackage{colortbl}
\usepackage{arydshln}

\usepackage{xcolor}  

\newcommand{\blue}[1]{\textcolor{blue}{#1}}

\newcommand{\rev}[1]{\marginpar{\raggedright\textcolor{blue}{#1}}}
\newcommand{\ms}[1]{\textcolor{black}{#1}}

\newcommand{\tonote}[1]{\textsuperscript{#1}}

\renewcommand{\blue}[1]{#1}

\renewcommand{\rev}[1]{}

\usepackage{listings}
\usepackage{xparse}

\NewDocumentCommand{\mintinline}{O{} m m}{%
  \texttt{\detokenize{#3}}%
}

% ...
\lstnewenvironment{minted}[2][]%
  {\lstset{
    basicstyle=\ttfamily\footnotesize,
    breaklines=true,
    columns=fullflexible,
    keepspaces=true,
    showstringspaces=false,
    frame=single
  }}%
  {}

\begin{document}

\title{GS-QA: A Benchmark for Geospatial Question Answering}

\author{Majid Saeedan}
\email{msaee007@ucr.edu}
\affiliation{%
  \institution{University of California, Riverside}
  \city{Riverside}
  \state{California}
  \country{USA}
  \postcode{92521}
}

\author{Muhammad Shihab Rashid}
\email{mrash013@ucr.edu}
\affiliation{%
  \institution{University of California, Riverside}
  \city{Riverside}
  \state{California}
  \country{USA}
  \postcode{92521}
}

\author{Ahmed Eldawy}
\email{eldawy@ucr.edu}
\affiliation{%
  \institution{University of California, Riverside}
  \city{Riverside}
  \state{California}
  \country{USA}
  \postcode{92521}
}

\author{Vagelis Hristidis}
\email{vagelis@cs.ucr.edu}
\affiliation{%
  \institution{University of California, Riverside}
  \city{Riverside}
  \state{California}
  \country{USA}
  \postcode{92521}
}


\begin{abstract}

Recent advances in Large Language Models (LLMs) have led to dramatic improvements in question answering (QA). To address the challenge of evaluating QA systems, standardized benchmarks have been introduced. This work focuses on the problem of geospatial QA, where a large collection of geospatial data is available in the form of a spatial database or other forms. Existing work on geospatial QA benchmarks has various limitations, including a small number of questions, limited spatial predicates, narrow output types, and no multi-source reasoning. We present GS-QA, an extensible geospatial QA benchmark with 2,800 question-answer pairs across 28 templates on top of OpenStreetMap and Wikipedia data, covering a wide range of spatial objects, predicates (including directional and towards filtering), and answer types (entity names, locations, distances, directions, counts, and aggregated areas/lengths). A key feature of GS-QA is that some questions require combining information from multiple sources, e.g., geospatial information from OSM and factual information from Wikipedia. GS-QA includes a comprehensive evaluation methodology that combines text-based QA measures with geospatial-specific measures such as distance error and angular error. We implemented nine LLM-based geospatial QA baselines using three LLMs (GPT-4o, Claude Sonnet 4.6, and Ministral-3) with combinations of direct prompting, retrieval-augmented generation, and text-to-SQL. Our results show that existing solutions perform reasonably well on simple spatial predicates with entity name outputs, but accuracy degrades significantly for questions involving complex spatial predicates, numeric output types, and multi-source reasoning, demonstrating that geospatial QA remains a challenging open problem warranting further research.
\end{abstract}

\maketitle



\section{Introduction}

Question Answering (QA) systems are designed to answer free-form questions. 
Earlier QA works  focused on questions based on a given text passage \cite{quac}, while more recent work focuses on open-domain QA, where a large collection of documents or other data must be searched to find the answer.
Recent advancements have shown that Large Language Models (LLMs) are excellent at synthesizing text responses to complex questions and can be used to build more robust QA systems \blue{\cite{devlin2019bert,raffel2020t5,brown2020gpt3}}\rev{R1.1}. Recent work has proposed benchmarks for the evaluation of the performance of LLMs in QA systems~\citeN{auer_sciqa_2023, krithara_bioasq-qa_2023, chen-etal-2023-theoremqa, 10.1145/3661304.3661901}. \blue{These benchmarks were proposed as LLM-based QA systems have become prevalent to evaluate capabilities enabled by generative models, such as open-ended answer synthesis and complex multi-step reasoning.} \rev{R2.1}

\begin{figure}
    \centering
    \includegraphics[width=.6\linewidth]{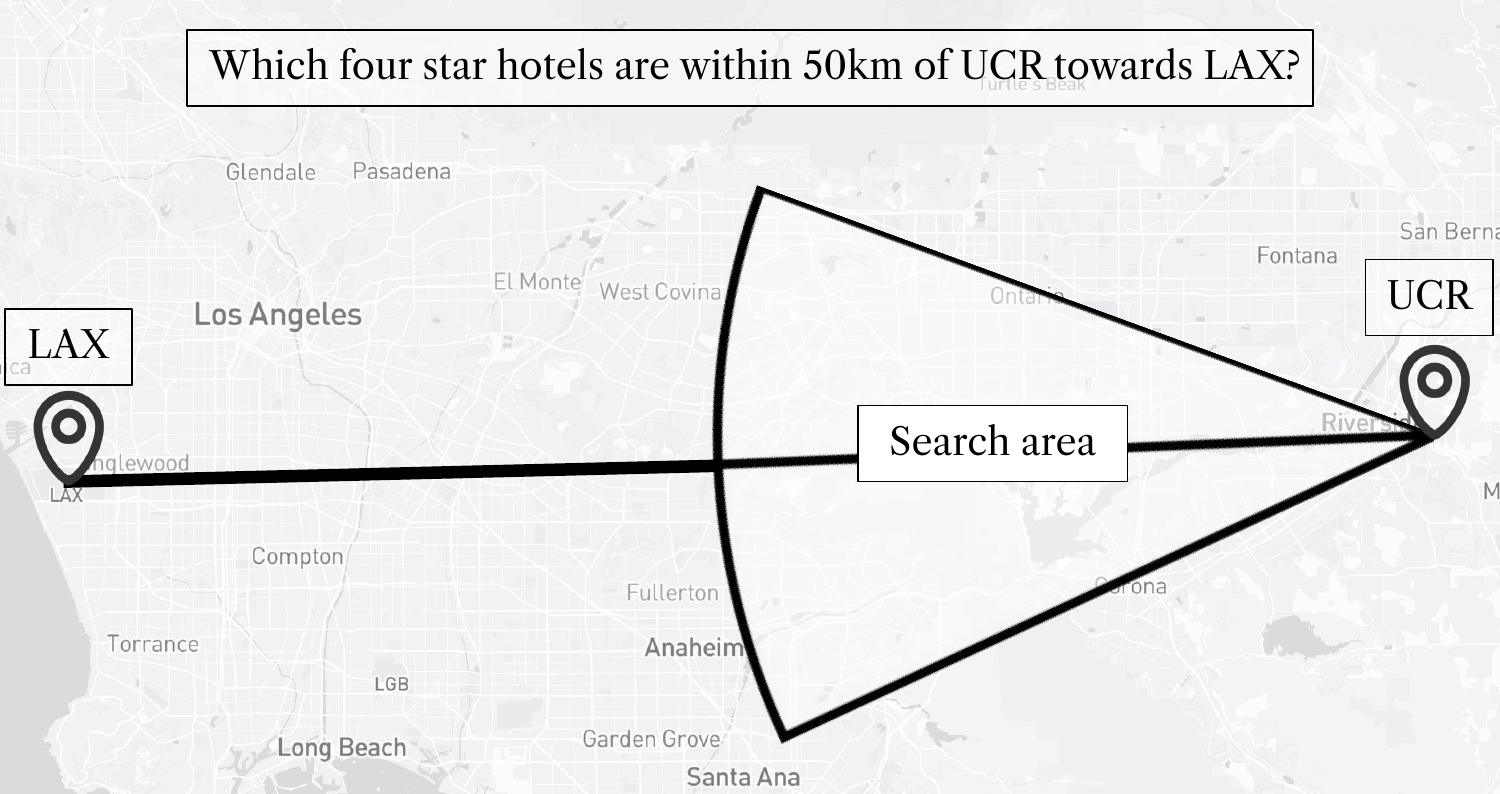}
    \caption{Geospatial Question Answering Example}
    \label{fig:intro_example}
\end{figure}

In the area of geospatial data management, QA has the potential to disrupt the way that people look for geospatial information, given the complexity of querying geospatial data for non-experts. 
As an example, consider the question \mintinline[fontsize=\footnotesize,breaklines]{python}{'Which four star hotels are within 50km of UCR towards LAX?'} 
To answer such a question, first, the anchoring locations must be identified, which are the Univesity of California, Riverside (\mintinline[fontsize=\footnotesize,breaklines]{python}{'UCR'}) and Los Angles International Airport (\mintinline[fontsize=\footnotesize,breaklines]{python}{'LAX'}). Then, their location coordinates must be retrieved. After that, the spatial predicates must be identified, which are a range query within a 50 kilometers radius and a direction filter based on an angle. Finally, hotels that are in locations that satisfy the spatial predicates are retrieved, as shown in Figure~\ref{fig:intro_example}.

\blue{Consistent with prior findings on the limitations of LLMs in spatial reasoning~\cite{ji2025foundation,kefalidis_question_2024}, state-of-the-art conversational AI tools like GPT-4 are unable to correctly answer such questions, because they do not account for all the spatial predicates.} \rev{R2.2}
For example, passing the example question to ChatGPT returns popular hotels in Riverside and some nearby cities that are not necessarily in the search area; that is, the directional predicate is ignored. Changing the question by replacing LAX with Las Vegas, which is towards the opposite direction, still produces the same answer.


Designing a system for answering such questions is challenging. 
First, understanding spatial predicates and performing spatial reasoning are required.
Current LLMs often respond with popular entities, e.g., the names of popular hotels, which may not necessarily be the best answer.
Second, the answers to some questions can change over time, so the LLM may not be up-to-date. 
Third, some questions may require synthesizing data from multiple sources, which may include unstructured data sources. 
Building an effective geospatial QA system could enable users and practitioners from many fields to get reliable answers without relying on specialized tools or languages.

There is limited work on benchmarking Geospatial Question Answering (GeoQA) \cite{kefalidis_question_2024,punjani_template-based_2018,karalis2019extending}\blue{; more recently, MapQA~\cite{li2025mapqa} introduced a closely related benchmark}. \rev{R2.3}
These have various limitations.
First, \blue{the crowdsourced benchmarks (GeoQuestions201, GeoQuestions1089)} include a small number of questions and no mechanism to generate more questions of given types. Second, they assume the existence of a spatial knowledge graph\blue{, which limits the query paradigm to GeoSPARQL}. Finally, they have limited spatial operators or non-spatial conditions. \rev{R2.4, R2.5}

Evaluating GeoQA is challenging for various reasons.
Geospatial data represent many different types of entities, such as points representing restaurants, lines for roads, polygons for region borders, etc. There are many types of questions that can be asked about this type of data, such as asking about directions, the location of an object that satisfies some conditions, and more, including questions that require aggregation and analysis of a large number of records. 
Further, the generated question-answer pairs must be unambiguous. A question must have a unique answer that can be computed deterministically given reference data. In addition, we need a clear process for evaluating the correctness and quality of the generated question-answer pairs. Furthermore, we want a benchmark that can be easily updated or expanded given reference data.

In this work, we introduce GS-QA, a benchmark for evaluating open-ended answers on questions that involve geospatial data. It includes 28 question templates, incorporating a variety of spatial objects, spatial predicates, and output types, among others. It also includes multi-source (i.e., multi-source) questions that require multiple steps to be answered, synthesizing information from multiple spatial and non-spatial sources. We provide one hundred questions for each template, for a total of 2800 questions. 
We also propose various evaluation strategies appropriate for GeoQA, which go beyond traditional text-based matching used in existing QA work. We complement text-based matching with spatial-based measures, such as relative distance error. 
We created and complemented a suite of six diverse LLM-based GeoQA baselines, which combine LLMs, retrieval, and structured querying (i.e., text-to-SQL). 

In summary, our contributions are as follows:
\begin{itemize}
    \item We create a GS-QA, a benchmark dataset comprising 2800 questions from 28 templates, with a wide range of spatial objects, predicates, and non-spatial information. We have published GS-QA along with the baselines' results \cite{qa_2025}.
    \item We develop a methodology for generating a large number of question-answer pairs based on our templated and a reference database, using corresponding SQL query templates.
    \item We implement several LLM-based GeoQA baselines, based on state-of-the-art retrieval methods.
    \item We propose a suite of text-based and geospatial-specific evaluation measures.
    \item We perform extensive experiments that show that existing baselines have low performance for most of the query templates of GS-QA.
\end{itemize}


We start by summarizing the related work in the literature and how GS-QA fills a needed gap in Section~\ref{sec:related}. In Section~\ref{sec:benchmark}, we present our methodology used to create the benchmark. In Section~\ref{sec:baselines}, we discuss the baselines that we use to demonstrate how to use the benchmark. In Section~\ref{sec:experiments}, we provide  experimental results evaluating the baselines. Finally, in Section~\ref{sec:future}  we discuss future work and conclude.




\section{Related Work}
\label{sec:related}

{\bf GeoQA Datasets.}

\blue{Early efforts in geographic question answering benchmarks included several works. GeoCLEF~\cite{mandl2008geoclef} evaluated cross-language geographic information retrieval from text document collections. GeoQuery~\cite{zelle1996geoquery}, a benchmark of 880 questions about U.S.\ geography was designed for semantic parsing. Another work is the GikiCLEF tasks~\cite{giki2022dataset}, which tested spatial reasoning over Wikipedia content. Moreover, GeoAnQu~\cite{beydokhti2021geoanqu} introduced a corpus of 429 geo-analytical questions compiled from GIS textbooks and scientific articles, targeting analytical workflows rather than factual retrieval.} \rev{R1.2}


\blue{More recent benchmarks for open-domain GeoQA include} GeoQuestions201~\cite{punjani_template-based_2018} and GeoQuestions1089~\blue{\cite{kefalidis2023benchmarking}}. \blue{GeoQuestions201 introduced template-based question answering over linked geospatial data using GeoSPARQL, and was developed through crowdsourcing. GeoQuestions1089 extended this effort with 1089 questions over the YAGO2geo knowledge graph~\cite{karalis2019extending}, providing both natural language questions and their corresponding GeoSPARQL queries.} \rev{R2.7}
\blue{The concurrent work MapQA~\cite{li2025mapqa} is closely related to ours: it introduces 3,154 question-answer pairs over OpenStreetMap for two U.S.\ regions (Southern California and Illinois), also using SQL query templates to generate QA pairs from a PostGIS database. MapQA covers nine question types including neighborhood inference and geo-entity type identification.} \rev{R2.3, R2.4}

\ms{We propose GS-QA \blue{as a complementary benchmark that addresses specific gaps in the existing landscape.} \blue{Both GS-QA and the knowledge-graph based benchmarks require the existence of organized geospatial data; the key difference is in the query paradigm and the spatial operations supported. The knowledge-graph based benchmarks (GeoQuestions201, GeoQuestions1089), GS-QA operates over a heterogeneous combination of a structured spatial database and external document sources (Wikipedia), adopting an open-retrieval approach.} \blue{Compared to MapQA, GS-QA covers a wider range of spatial predicates, including directional filtering and the the \emph{towards}. It also includes multi-source questions that require synthesizing information from both the spatial database and Wikipedia. Furthermore, GS-QA gives more focus to output type and has more output types as compared to the predominantly text based answers in MapQA.}} \rev{R2.5, R2.14, R2.16}
Second, all GS-QA questions reference multiple entities, including the question itself and when computing the answer. This is to ensure that the system is evaluated based on performing spatial operations and not simple retrieval based on string matching.
Furthermore, our proposed benchmark includes a larger number of question categories and each category has a larger variety of questions. Also, in being able to automate the question generation process, our benchmark can be easily extended to include more question categories.
These differences are summarized in Table~\ref{tab:benchmark_comparison}.

\blue{It is worth noting that MapQA follows a similar template-based SQL generation methodology to ours, while GeoQuestions201 and GeoQuestions1089 relied on crowdsourcing to collect natural language questions. Our generation methodology differs in that each template is paired with a parameterized SQL query that both instantiates the question and deterministically computes the answer, enabling automatic answer verification and benchmark updating as the underlying OSM data evolves (Section~\ref{sec:update}). This approach is inspired by template-based benchmark generation in the broader Text2SQL literature~\cite{zhongSeq2SQL2017,li_can_2023}, adapted here for the geospatial domain.} \rev{R2.4, R2.6, R2.11}

%
%





\begin{table*}[h]
\centering
\caption{\blue{Comparison between GS-QA and existing GeoQA benchmarks [R2.14, R2.16]}}
\label{tab:benchmark_comparison}

\footnotesize
\begin{tabular}{|l|l|l|l|l|l|}

\hline
\textbf{} & \textbf{GS-QA} & \textbf{GeoQuestions1089} & \textbf{GeoQuestions201} & \textbf{MapQA} & \textbf{GeoAnQu} \\
\hline
Main reference data & OSM & YAGO2geo KG & Linked geospatial data & OSM & GIS textbooks \\
\hline
Query language & SQL (PostGIS) & GeoSPARQL & GeoSPARQL & SQL (PostGIS) & GIS workflows \\
\hline
Multi-source retrieval & Yes & No & No & No & No \\
\hline
Question categories & 28 & 9 & 7 & 9 & N/A \\
\hline
Number of questions & 2800 & 1089 & 201 & 3154 & 429 \\
\hline
Creation method & Templates & Crowdsourced & Crowdsourced & Templates & Manual \\
\hline
Easily extensible & Yes & No & No & Yes & No \\
\hline
Geographic scope & USA & Global (KG) & Global (KG) & USA (2 regions) & N/A \\
\hline
\multicolumn{6}{|l|}{\textbf{Spatial Concepts Covered}} \\
\hline
Nearest neighbor & \checkmark & \checkmark & \checkmark & \checkmark &  \\
\hline
Range & \checkmark &  & \checkmark & \checkmark &  \\
\hline
Direction & \checkmark &  &  & \checkmark &  \\
\hline
Towards & \checkmark &  &  &  &  \\
\hline
Region containment & \checkmark & \checkmark & \checkmark &  &  \\
\hline
Topological &  & \checkmark & \checkmark &  &  \\
\hline
Buffer &  & \checkmark &  &  &  \\
\hline
Adjacency / closeness &  &  &  & \checkmark &  \\
\hline
Distance computation & \checkmark &  &  & \checkmark &  \\
\hline
Proximity comparison &  &  &  & \checkmark &  \\
\hline
Aggregation & \checkmark & \checkmark &  &  &  \\
\hline
Non-spatial attribute filtering & \checkmark &  &  & \checkmark &  \\
\hline
\multicolumn{6}{|l|}{\textbf{Output Types}} \\
\hline
Entity name & \checkmark & \checkmark & \checkmark & \checkmark & \\
\hline
Location (coordinates / address) & \checkmark &  & \checkmark &  &  \\
\hline
Direction (angle / compass) & \checkmark &  &  &  &  \\
\hline
Distance (numeric) & \checkmark &  &  & \checkmark &  \\
\hline
Count & \checkmark & \checkmark &  &  &  \\
\hline
Area / Length (aggregated) & \checkmark &  &  &  &  \\
\hline
Attribute value & & \checkmark &  & \checkmark &  \\
\hline
External (out-of-schema) & \checkmark &  &  &  &  \\
\hline
\end{tabular}
\end{table*}

{\bf GeoQA Systems.} \blue{A range of systems have been developed for geospatial question answering, employing diverse approaches. Early systems such as Chen et al.~\cite{chen_synergistic_2013} used frameworks combining information retrieval with spatial reasoning. Li et al.~\cite{li_neural_2021} proposed neural factoid geospatial QA using dependency parsing and knowledge graph querying. The GeoQA2 system~\cite{kefalidis_question_2024} extended it with improved template-based parsing and materialized topological relations for more efficient GeoSPARQL query execution.} \rev{R2.8}

\blue{More recent systems leverage LLMs directly. 
GeoQAMap~\cite{feng2023geoqamap} integrates LLMs with the Wikidata knowledge base, translating natural language questions into SPARQL queries and visualizing results as interactive maps. Spatial-RAG~\cite{yu2025spatialrag} proposes a spatial retrieval-augmented generation framework specifically designed for real-world spatial reasoning questions, addressing the limitation that standard RAG methods do not account for spatial proximity in their retrieval step. Our baselines are broadly representative of these system categories, bare LLM, Text2SQL, and RAG. Our benchmark is designed to be system-agnostic, enabling evaluation of any GeoQA architecture.} \rev{R2.8}
\blue{Beyond factoid GeoQA, there is a growing body of work on LLM-powered natural language interfaces for GIS more broadly, including autonomous GIS agents such as LLM-Geo~\cite{li2023autonomous} that translate natural language into executable GIS workflows, and GeoGPT~\cite{zhang2024geogpt} that uses LLMs as the reasoning core for spatial analysis tasks. While these systems target general GIS automation rather than factoid QA specifically, they represent part of the broader landscape that GS-QA could help evaluate. We also note that our benchmark is designed to evaluate more basic functionality, such as identifying spatial entities, directions, and other spatial information correctly. We find that it is still challenging for LLMs to identify them and map them correctly to queries or generate embeddings that map them correctly. As we note Section~\ref{sec:discussion}, we leave evaluation of more advanced spatial analysis as future work.
}\rev{R2.8}


{\bf Closed-domain Spatial Reasoning.}
There has been work on evaluating spatial reasoning, such as SPARTQA \cite{mirzaee2021spartqa}. The problem here is: given a description (e.g. ``We have three blocks, A, B, and C. Block B is to the right of block C and it is below block A.") we ask a question (e.g. ``Which object is above a medium black square?") that requires spatial reasoning on the provided description. This work is different from our problem, which is open-domain question answering.
There is also work on visual spatial reasoning \cite{liu2023visual}, where the input is an image instead of a textual description. Similar works also fall under the domain of spatial reasoning and spatial proximity, including \cite{chen_geoqa_2021, li_proximity_2024}. The work in~\cite{kazemi_beydokhti_probabilistic_2024} evaluates spatial reasoning in the context of GeoQA.

\blue{Additional relevant work on evaluating LLMs' spatial reasoning abilities includes~\cite{beydokhti2024integrating}, that investigated integrating LLMs with qualitative spatial reasoning frameworks, and~\cite{cohn2023dialectical}, that provided an initial appraisal of LLMs commonsense spatial reasoning capabilities. Furthermore, \cite{ji2025foundation} conducted a systematic assessment of LLMs ability to understand geometries and topological spatial relations, finding significant limitations. These findings are consistent with the low baseline performance we observe on GS-QA, particularly for questions involving directional and distance-based predicates.} \rev{R1.2}


{\bf Benchmarks for LLM-based QA Evaluation.}
\blue{With the rapid and wide spread of LLMs, researchers have developed several benchmarks to evaluate QA systems, many of which test capabilities particularly relevant to LLM-based approaches.} \rev{R2.1}
Some benchmarks rely on multiple choice questions, e.g., MMLU~\cite{hendryckstest2021} and GPQA~\cite{rein2024gpqa}. These benchmarks are more challenging, and they require deep domain knowledge. The proposed benchmark has more open-ended answers, and we take that into account when designing the evaluation strategy, as detailed in the paper.


There are also domain-specific benchmarks such as MATH-500~\cite{lightman2023lets} which evaluates LLMs on solving math problems in a step-by-step approach. In this case, all the steps are used to evaluate the model and not just the final answer. Our benchmark focuses on spatial QA and it evaluates only the final answer. We believe that step-by-step evaluation can be useful but we leave this for future work. In this case, the answer will need to change to steps that resemble how GIS analysts approach the question. Another difference is that the MATH-500 dataset was used for both training and evaluation while this paper focuses on evaluation. Our QA dataset generator can also be used to generate data for training but we leave this also for future work.


The evaluation of LLMs is very challenging, and the process is not yet standardized. The work in~\cite{laskar-etal-2024-systematic} discusses the challenges and strategies for LLM evaluation. Some of the challenges include contamination of benchmark data in training, tailoring prompts to include examples very similar to the questions being asked, and reproducibility issues. This informs how we designed our baselines. \rev{R2.9}

{\bf Multi-\blue{source} questions.} \rev{R2.6}
\ms{These types of questions require multiple steps in order to be answered, because they typically involve fact-chaining. There are existing datasets for these types of questions including \cite{yang2018hotpotqa,talmor2018webquestions,ho2020twiki,trivedi2022musique}, and \cite{qiu2022survey} provides a survey on this topic. The main difference in our benchmark is that all the facts are related to a spatial object, either identifying a spatial object by its that is referenced in the question or first answering a question by retrieving a spatial object and then retrieving an external fact related to it. \blue{We use the term ``multi-source'' rather than ``multi-hop'' throughout this paper, since our questions specifically require synthesizing information from multiple data sources (the spatial database and Wikipedia), which is a more precise characterization than the general notion of multi-hop reasoning within a single source.}} \rev{R2.6}

{\bf Text2SQL.}
Converting a natural language question to an SQL query is another active area of research. In the literature, this problem is usually referred to as Text2SQL or NL2SQL.
Several methods have been proposed, including LLM-based approaches \cite{li2024dawn}.
There exist several benchmarks for this problem, such as \cite{zhongSeq2SQL2017,li2024can}, but they are not focused on geospatial queries.

\blue{Recent work has begun addressing Text2SQL specifically for geospatial data. Monkuu was proposed in~\cite{yu2025monkuu}, which is an LLM-powered natural language interface for geospatial databases that employs dynamic schema mapping to handle the complexity of spatial data structures. In~\cite{wang2025gptspatialsql}, they explored GPT-based Text2SQL specifically for spatial databases, evaluating how well LLMs can generate queries involving spatial functions. Commercial tools such as Kinetica's SQLGPT~\cite{kinetica_sqlgpt} also target natural language interfaces for spatial data. MapQA~\cite{li2025mapqa} evaluates Text2SQL as one of its baseline approaches, finding that LLMs perform well on single-hop spatial queries but struggle with multi-hop reasoning. Our benchmark complements these efforts by providing a broader set of spatial predicates (including direction and towards filtering) that can serve as a challenging evaluation suite for geospatial Text2SQL systems. We did not adopt these existing systems as baselines because our goal was to evaluate general-purpose LLM approaches rather than specialized geospatial NLIs, but future work could use GS-QA to benchmark such systems.} \rev{R1.2, R2.11}

\section{GS-QA Benchmark Creation}
\label{sec:benchmark}

This section outlines our methodology for creating the GS-QA benchmark's reference data, questions, and answers.
First, Section~\ref{sec:ref_data} explains how we extract and prepare geospatial \textbf{reference data} from OpenStreetMap.
Then, Section~\ref{sec:templates} discusses the creation of \textbf{question templates} by selecting predicates, spatial entities, and answer types. 
After that, Section~\ref{sec:questionsgen} describes how to \textbf{instantiate these templates} to generate questions. 
Section~\ref{sec:quality} explains the process of \textbf{verifying question quality} and answer correctness to ensure question diversity.
Finally, Section~\ref{sec:update} discusses how to \textbf{keep the benchmark up-to-date} as the reference data is updated.

\subsection{Reference Database}\label{sec:ref_data}

The first step in creating the benchmark is to prepare the reference dataset that the LLM will use to answer all questions. Our goal is to create a large number of questions that are primarily focused on spatial objects and spatial operations. However, there are four challenges in doing that, as mentioned earlier in the introduction. These include ensuring the questions have a large variety, there is a standardized method to get the correct answers, scalability, and handling data updates. The reference database we create is a structured database instance that we generate from OpenStreetMap (OSM). Building a reference database with a known structure helps us with these challenges.
First, it provides us with a clear structure about the type of spatial objects available and the attributes associated with them. This makes it more straightforward to create a diverse set of question templates.
Secondly, the support of standardized SQL queries by the database makes it possible to create queries that instantiate the question templates, as well as define their answers.
Furthermore, by having standardized queries, it is possible to generate question/answer pairs on a scale.
Additionally, there are also advantages in relation to data updates and handling data from multiple sources in the future. The database can also serve as a reference in RAG (Retrieval Augmented Generation) systems.


We get our source data from OSM. We use it as our source because it is the largest publicly available geospatial dataset. However, our methodology can be used on any source data. We don't use OSM directly since it is in a semi-structured format, and converting it to a relational database provides all the benefits mentioned earlier. \blue{We chose to organize our reference data as a relational database with PostGIS rather than leveraging an existing knowledge graph such as WorldKG~\cite{dsouza2021worldkg}. The choice of a relation structured database was motivated by three practical considerations. First, SQL with PostGIS is the dominant query language in GIS practice, and a relational representation allows the benchmark to serve dual purposes: evaluating QA systems and evaluating geospatial Text2SQL approaches. Second, PostGIS provides direct access to a rich set of spatial operations (e.g., \texttt{ST\_Azimuth}, \texttt{ST\_Intersects}) that map naturally to our spatial predicates; some of these operations, particularly azimuth-based directional filtering and the towards predicate, do not have direct equivalents in GeoSPARQL or typical KG query languages. Third, the relational representation enables deterministic answer computation through SQL, supporting automatic benchmark generation and updating. We did not use WorldKG specifically because, while it provides a semantic representation of OSM data, its schema is designed around RDF triples and GeoSPARQL queries, which would not support the full range of spatial predicates in our benchmark (e.g., directional filtering via azimuth computation). However, a KG-based variant of GS-QA using WorldKG or a similar resource could be a valuable complement and is a direction for future work.} \rev{R2.5}

To build the reference database, we first obtain the most recent version of OpenStreetMap for the entire United States of America from GeoFabrik OpenStreetMap Extracts~\cite{noauthor_geofabrik_nodate}, which are updated daily. We use the extract produced on February 2, 2024. We show in Section~\ref{sec:update} how to update the generated question-answer pairs using more recent versions.

Then, we use OSMX~\cite{singla_osmx_2022} to extract five different datasets.
1)~\textbf{Points of interest (POI)}, which includes points for restaurants, shops, hospitals, and many more.
2)~\textbf{Administrative boundaries}, which contains boundaries for states, counties, cities, etc., at different administrative levels.
3)~\textbf{Parks} contains boundaries for recreational parks, nature reserves, and sports stadiums, among others.
4)~\textbf{Water bodies} which includes lakes, rivers, streams, and others.
5)~\textbf{Roads and walkways} includes roads for cars and pedestrians.
Readers can refer to~\cite{singla_osmx_2022} for more details on the extraction process.


Each object in OSM has a geometry attribute, which is the main attribute relevant to creating our questions. Additionally, each object has a set of tags which add more information to each object, e.g., park name or speed limit. Since tags are user-defined, they can be very arbitrary and it would be difficult to store all of them as separate attributes. Instead, we identified some common attributes that are useful for question answering, such as name, a reference to a Wikipedia page, and address. Then, we selected a small subset of other attributes that are related to specific categories like points of interest that provide services, or are related to tourism, etc. These columns provide us with enough variety to create a large set of natural questions. 
We summarize the selected attributes for all tables in the database in \autoref{tab:database_summary}.
Refer to~\cite{noauthor_map_nodate} for details about different attributes and their possible values.
Using the schema (columns) in \autoref{tab:database_summary}, we create a database instance using PostgreSQL with PostGIS. The statistics of our reference database are shown in~\autoref{tab:database_summary}.

 \begin{table}[]
\centering
\caption{Reference database summary}
\label{tab:database_summary}
\footnotesize
\begin{tabular}{|l|l|l|l|}
\hline
\textbf{Table} & \textbf{Columns/Attributes}                                                                              & \textbf{Geometries} & \textbf{Records} \\ \hline
\textbf{POI} &
  \begin{tabular}[c]{@{}l@{}}Geometry, OSM ID, Name,\\ Wikipedia, Address, Leisure,\\ Amenity, Tourism, Emergency,\\ Restaurant attributes.\end{tabular} &
  Points &
  267612 \\ \hline
\textbf{Park}  & \begin{tabular}[c]{@{}l@{}}Geometry, OSM ID, Name,\\ Wikipedia, Address, Leisure.\end{tabular}           & All types           & 5997948          \\ \hline
\textbf{Lake}  & \begin{tabular}[c]{@{}l@{}}Geometry, OSM ID, Name,\\ Wikipedia, Address, Water,\\ Waterway.\end{tabular} & All types           & 7988851          \\ \hline
\textbf{Road}  & \begin{tabular}[c]{@{}l@{}}Geometry, OSM ID, Name,\\ Wikipedia, Address, Highway.\end{tabular}           & All types           & 36827649         \\ \hline
\textbf{Region} &
  \begin{tabular}[c]{@{}l@{}}Geometry, OSM ID, Name,\\ Wikipedia ID, Address,\\ Border type,\\ Adminstration level.\end{tabular} &
  \begin{tabular}[c]{@{}l@{}}LineStrings\\ Polygon\end{tabular} &
  39137 \\ \hline
\end{tabular}
\end{table}

The name attribute is used to refer to entities in our questions-answer pairs. The columns related to Wikipedia are used to retrieve information related to question entities to create multi-source questions that contain 
information from an additional information source. This can provide another dimension for evaluating question answering systems. 

The addresses are stored in multiple columns and are used for disambiguation, since an object's name may exist in many locations. Some address attributes like state and county are appended to names to make them less ambiguous. The address attributes are also used as an additional indicator of the importance of a POI, since many records do not have an address defined, although it is relatively straightforward to translate the coordinates to an address, but it may not always be accurate.

The {\em amenity} column is associated with points of interest that provide everyday services, like restaurants and coffee shops. Although there are many types of amenities, we selected only a few types that would create natural-sounding questions and still provide a wide variety of values. We only considered amenities related to sustenance such as restaurants and coffee shops, as well as hospitals and universities when we created the questions. When the value for this column is restaurant, other columns can also be used to add additional non-spatial descriptors such as the type of cuisine, availability of drive-through, or outdoor seating. Later, this helps to add more variety to the questions.

From the tourism column, we selected ten categories, including hotels, museums, theme parks, art galleries, etc. The museum column is used to add a non-spatial attribute to the questions when the value in the tourism column is a museum.

In the leisure column, we consider values like park, beach resort, golf course, nature reserve, sports center, among others. For water bodies, we consider lakes, bays, rivers, streams, and a few others. For roads, we consider primary, secondary, residential, and more.


\subsection{Question Templates}\label{sec:templates}
To create a large number of questions that can efficiently benchmark a spatial QA system, we first create {\em question templates}, and then use these templates to instantiate an arbitrarily large number of questions. \autoref{tab:question_templates} lists all the 28 templates that we use. Each template has one or more {\em variables} or placeholders that are substituted from the database to create a question instance. \autoref{tab:spatial-objects} lists all these variables and how they get substituted. For example, template {\bf T1} can be instantiated into the question `{\em Can you suggest a {\bf restaurant} within {\bf 1 km} from {\bf San Diego Zoo}?}'.

A question template is defined by (a) one or more predicates (e.g., ``nearest neighbor''), (b) one or more spatial anchor entities (e.g., ``Yosemite National Park''), (c) output type (e.g., ``location''), and (d) a set of text phrases that can be used to instantiate a question.

\blue{The 28 templates were derived through a systematic combinatorial process. We enumerated combinations of the five spatial predicates (nearest neighbor, range, direction, towards, intersects) with the eight output types (entity name, location, direction, distance, count, area, length, external). From this combinatorial space, we selected the subset of combinations that produce natural-sounding questions an everyday user might ask. Some combinations were excluded because they would produce unnatural phrasing (e.g., asking for the ``area'' of a nearest-neighbor POI). We then augmented the selected combinations with non-spatial predicate variants (T2, T6, T14, T18) and multi-source variants (T7, T8). Each template was manually reviewed for naturalness and unambiguity. We also avoided including more complex questions that that are more likely to be asked by geospatial data analysts, since we wanted to focus on more general use cases, and we found that the questions that templates we included are already challenging for existing LLMs.} \rev{R1.3, R2.11}


Additionally, for each question template, we create an associated SQL query template that computes the answer to the question when applied to the reference database. 

The templates are based on a variety of combinations of predicates and output types. The table shows only one text example for each template, but each template has several text phrases to choose from, adding to the richness of the generated questions. \ms{They are based on a subset of spatial predicates, metrics, and aggregate functions from all the operations supported by PostGIS. It would be intractable to include all combinations of operations and functions of PostGIS in the benchmark. Instead, we focused on a smaller subset of combinations that can be easily translated to natural language questions that can be asked by everyday users. We note that even this smaller subset is already very challenging to answer using existing LLM based tools, as will be shown later. However, based on the question generation process that we define here. The benchmark can be easily extended to support more operations and variety of questions in later iterations.
} 

Furthermore, for some of these templates, specifically T1, T5, T13, and T17, we created four more templates that include an additional non-spatial predicate, specifically T2, T6, T14 and T18, respectively. These are based on some of the attributes we included in the reference database, such as the type of cuisine, the type of museum, etc. Additionally, we created two more templates from T5 as  multi-source questions, which are T7 and T8 that include information that does not exist in the reference database. For T7, the information the question is asking about does not exist in the reference database. Alternatively, for T8, the name of the anchor point in the question is replaced by unique information about it retrieved from another source. Both types of questions add another layer of difficulty in which the QA system must synthesize information from multiple sources to find the final answer. Next, we cover the building blocks of these templates in more detail.

\blue{We note that some templates share the same spatial predicates but differ in their expected output type. For example, T1 (``Can you suggest...'') expects an entity name (e.g., ``Hilton Garden Inn''), while T13 (``Where can I find...'') expects a location (e.g., coordinates or an address). This distinction is important for evaluation: a system may correctly identify an entity but fail to provide its location, or vice versa. The different output types require different evaluation metrics which we discuss in Section~\ref{sec:eval}, which is why they are treated as separate templates.} \rev{R1.4}

\begin{table*}[]
\caption{Question templates}
\label{tab:question_templates}
\resizebox{\textwidth}{!}{%
\begin{threeparttable}

\begin{tabular}{|l|l|l|l|}
\hline
\textbf{ID}  & \textbf{Text}  & \textbf{Output Type} & \textbf{Spatial Predicates}  \\ \hline

T1  & Can you suggest \{POI\_CAT\} within \{DISTANCE\} from \{ANCH\_POI\}?  & \multirow{12}{*}{Entity name}  & Range  \\ \cline{1-2} \cline{4-4}
T2\tonote{*}  & Can you suggest \{POI\_NONSPAT\} within \{DISTANCE\} from \{ANCH\_POI\}?  &   & Range  \\ \cline{1-2} \cline{4-4}
T3  & Which \{POI\_CAT\} is located within \{DISTANCE\} in the \{DIRECTION\} of \{ANCH\_POI\}?  &  & Range, Direction  \\ \cline{1-2} \cline{4-4}
T4  & Which \{POI\_CAT\} can I find within \{DISTANCE\} from \{ANCH\_POI1\} towards \{ANCH\_POI2\}?  &  & Range, Towards  \\ \cline{1-2} \cline{4-4}
T5  & What is the nearest \{POI\_CAT\} from \{ANCH\_POI\}?  &  & Nearest Neighbor  \\ \cline{1-2} \cline{4-4}
T6\tonote{*}  & What is the nearest \{POI\_NONSPAT\} from \{ANCH\_POI\}?  &  & Nearest Neighbor  \\ \cline{1-2} \cline{4-4}
T7\tonote{\P} & What is the capacity of the nearest \{POI\_CAT\} from \{ANCH\_POI\}? & & Nearest Neighbor \\ \cline{1-2} \cline{4-4}
T8\tonote{\ddag}  & What is the nearest \{POI\_CAT\} from \{ANCH\_POI.external\}?  &  & Nearest Neighbor  \\ \cline{1-2} \cline{4-4}
T9  & What is the closest \{POI\_CAT\} \{DIRECTION\} of \{ANCH\_POI\}?  &  & Nearest Neighbor, Direction  \\ \cline{1-2} \cline{4-4}
T10  & What is the closest \{POI\_CAT\} from \{ANCH\_POI1\} towards \{ANCH\_POI2\}?  &  & Nearest Neighbor, Towards  \\ \cline{1-2} \cline{4-4}
T11  & What is the largest \{PARK\_WATB\} in \{REGION\}?  &  & Intersects  \\ \hline
T12  & What is the longest \{ROAD\_WATW\} in \{REGION\}?  &  & Intersects  \\ \cline{1-2} \cline{4-4}
T13  & Where can I find \{POI\_CAT\} within \{DISTANCE\} from \{ANCH\_POI\}?  &  \multirow{8}{*}{Location}  & Range  \\ \cline{1-2} \cline{4-4}
T14\tonote{*}  & Where can I find \{POI\_CAT\} within \{DISTANCE\} from \{POI\_NONSPAT\}?  &   & Range  \\ \cline{1-2} \cline{4-4}
T15  & Where can I find \{POI\_CAT\} located within \{DISTANCE\} in the \{DIRECTION\} of \{ANCH\_POI\}?  &   & Range, Direction  \\ \cline{1-2} \cline{4-4}
T16  & What location has \{POI\_CAT\} within \{DISTANCE\} from \{ANCH\_POI1\} towards \{ANCH\_POI2\}?  &   & Range, Towards  \\ \cline{1-2} \cline{4-4}
T17  & Where can I find the nearest \{POI\_CAT\} from \{ANCH\_POI\}?  &   & Nearest Neighbor  \\ \cline{1-2} \cline{4-4}
T18\tonote{*}  & Where can I find the nearest \{POI\_NONSPAT\} from \{ANCH\_POI\}?  &   & Nearest Neighbor  \\ \cline{1-2} \cline{4-4}
T19  & Where is the closest \{POI\_CAT\} \{DIRECTION\} of \{ANCH\_POI\}?  &   & Nearest Neighbor, Direction \\ \cline{1-2} \cline{4-4}
T20  & Where is the closest \{POI\_CAT\} from \{ANCH\_POI1\} towards \{ANCH\_POI2\}?  &   & Nearest Neighbor, Towards  \\ \hline
T21  & In which direction is \{POI\_CAT\} located within \{DISTANCE\} from \{ANCH\_POI\}?  &  \multirow{2}{*}{Angle}  & Range  \\ \cline{1-2} \cline{4-4}
T22  & What is the direction towards the closest \{POI\_CAT\} from \{ANCH\_POI\}?  &   & Nearest Neighbor  \\ \hline
T23  & How many \{POI\_CAT\} within \{DISTANCE\} from \{ANCH\_POI\}?  &  \multirow{2}{*}{Count}  & Range  \\ \cline{1-2} \cline{4-4}
T24  & How many \{POI\_CAT\} are there in \{REGION\}?  &   & Intersects  \\ \hline
T25  & How far can I find \{POI\_CAT\} within \{DISTANCE\} from \{ANCH\_POI\}?  &  \multirow{2}{*}{Distance}  & Range  \\ \cline{1-2} \cline{4-4}
T26  & How far is the closest \{POI\_CAT\} from \{ANCH\_POI\}?  &   & Nearest Neighbor  \\ \hline
T27  & What is the total area of all \{PARK\_WATB\} in \{REGION\}?  & Area  & Intersects  \\ \hline
T28 & What is the total length of all \{ROAD\_WATW\} in \{REGION\}?  & Length  & Intersects  \\ \hline

\end{tabular}%

\begin{tablenotes}
\item[*] Includes an additional non-spatial predicate, where POI\_CAT is more specific.
\item[\P] A multi-source question that asks about information retrieved from another source.
\item[\ddag] A multi-source question where the ANCH\_POI is replaced with unique information from an external source.
\end{tablenotes}
\end{threeparttable}

}
\end{table*}

\begin{table}[]
\caption{Question Parameters Summary}
\label{tab:spatial-objects}
\footnotesize
\begin{tabular}{|l|l|}
\hline
Parameter   & Description                                                                                                                                                                                                                                           \\ \hline
ANCH\_POI  & \begin{tabular}[c]{@{}l@{}}Name of anchoring POI from one of these categories:\\ aquarium, attraction, viewpoint, art gallery, theme park,\\ museum, gallery, zoo, hotel, university, park, nature reserve, \\ garden, stadium, hospital\end{tabular} \\ \hline
POI\_CAT   & \begin{tabular}[c]{@{}l@{}}POI category name: restaurant, café, fast food,\\ plus all the categories in ANCH\_POI.\end{tabular}                                                                                                                       \\ \hline
POI\_NONSPAT   & \begin{tabular}[c]{@{}l@{}}A more specific POI category,\\like cuisine for a restaurant, or museum type.\end{tabular}                                                                                                                       \\ \hline
REGION     & \begin{tabular}[c]{@{}l@{}}Name of region from one of: city, town, village,\\island, municipality, county, neighborhood, suburb, state\end{tabular}                                                                                                  \\ \hline
ROAD\_WATW & \begin{tabular}[c]{@{}l@{}}Name of road or waterway type, including:\\ primary, residential, pedestrian, etc., for roads\\ river, stream, and others for waterways.\end{tabular}                                                                      \\ \hline
PARK\_WATB & \begin{tabular}[c]{@{}l@{}}Name of park or water body types, including: \\ nature reserve, park, garden, golf course, etc., for parks\\ lake, bay, etc., for water bodies\end{tabular}                                                                \\ \hline
DISTANCE   & Random distance in the range {[}1,200{]} kilometers.                                                                                                                                                                                                  \\ \hline
DIRECTION  & Direction one of: north, northeast, east, southeast, etc.                                                                                                                                                                                             \\ \hline
\end{tabular}%
\end{table}


\bigskip
\noindent\underline{Question Parameters}


\noindent\textbf{\textit{Anchor POI} (ANCH\_POI):} This refers to a POI that is used to anchor the location of the search for the desired entity. The categories we selected for this type are more prominent POIs like attractions, museums, universities, and others. We avoided points that can exist repeatedly within a small geographic area, such as restaurants. To further reduce ambiguity, we attach the city and state to the name of the anchor point in a question. \blue{This was done to ensure unambiguous ground truth for evaluation purposes; a robust QA system should ideally be able to disambiguate entities from context alone. Future versions could include ambiguous variants as additional challenge questions to test entity disambiguation capabilities.} \rev{R2.12}

\noindent\textbf{\textit{POI category} (POI\_CAT):} This refers to the type of POI the question asks about. The answer to the question includes a non-spatial filter that selects points from this category. Any type of POI category can be used, but we require that it be from a different category than the anchoring point \blue{as a design choice to maximize question diversity across the benchmark. We acknowledge that same-category queries are realistic (e.g., finding the nearest Italian restaurant when at another Italian restaurant) and could be included in future versions.} \rev{R2.13}

As an example, we show how the previous two parameters can be used to instantiate a template such as \mintinline[fontsize=\footnotesize,breaklines]{python}{'Where can I find the nearest {POI_CAT} from {ANCH_POI}?'} (T17 in \autoref{tab:question_templates}). The output category could be instantiated with \textit{restaurant} while the anchor point could be \textit{Alaska Pacific University, Anchorage, AK}. Note that it is easy to also use other types of objects such as roads as the anchoring object in a question, using the same text phrase and the same query that obtains the answer.

\noindent\textbf{\textit{More specific POI category} (POI\_NONSPAT):} This is similar to the previous variable, but it represents a more specific POI\_CAT. For example, instead of just instantiating with \textit{restaurant}, the question can be instantiated with something more specific like \textit{restaurant with outdoor seating}. This is discussed in more detail later in this section.

\noindent\textbf{\textit{Administrative regions} (REGIONS):} all the regions in the region table as described in \autoref{sec:ref_data} are used in questions involving topological operations and aggregations.

\noindent\textbf{\textit{Parks and water bodies} (PARK\_WATB):} The parks and water bodies, like lakes, are used to instantiate questions that ask about areas. An example of such a template is \mintinline[fontsize=\footnotesize,breaklines]{python}{'What is the largest {PARK_WATB} in {REGION}?'} (T11 in \autoref{tab:question_templates}).

\noindent\textbf{\textit{Roads and water ways} (ROAD\_WATW):} the different types of roads and waterways such as rivers are used to instantiate questions asking about length. For example, a template that includes these object is \mintinline[fontsize=\footnotesize,breaklines]{python}{'What is the total length of all {RD_WATW} in {REGION}?'} (T28 in \autoref{tab:question_templates}).

These parameters are summarized in \autoref{tab:spatial-objects}. These make up a major spatial component in the questions. The other major spatial component is the spatial filtering operations.

\bigskip
\noindent\underline{Spatial Predicates}

All the questions that we generate contain at least one spatial predicate. In addition, the predicates determine what types of spatial objects are used. Next, we define the five spatial predicates that we included.

\noindent\textbf{\textit{Nearest Neighbor}:} this is the simplest spatial predicate, and it filters based on spatial proximity. It requires an anchoring point, such as the user's location, and provides the nearest object that satisfies other predicates in the question. An SQL query for a question that contains this predicate will include the following: 

\begin{minted}
[
fontsize=\footnotesize
]{postgresql}
ORDER BY  geometry <-> anchor_point ASC LIMIT 1;
\end{minted}

\noindent\textbf{\textit{Range}:} similar to the nearest-neighbor predicate, but also requires providing a distance limit. In this predicate, there might be multiple possible answers within the provided distance. The SQL will include something like:

\begin{minted}
[
fontsize=\footnotesize
]{postgresql}
ST_DWithin(geometry, anchor_point, distance)
\end{minted}

\noindent\textbf{\textit{Direction}:} This adds another predicate to a question, which filters based on the direction the user is interested in, such as: north, west, northeast, etc. We define eight directions and specify a specific angle range for each as shown in Figure~\ref{fig:direction_circle}. \blue{The eight 45-degree sectors follow the standard eight-point compass rose, which is the most commonly used directional division in everyday language and GIS applications. We set $0°$ as true north, consistent with PostGIS's \texttt{ST\_Azimuth} function.} \rev{R1.6}
 For example, the northeast direction will have the following SQL predicate:

\begin{minted}
[
fontsize=\footnotesize
]{postgresql}
degrees(ST_Azimuth(anchoring_point, geometry)) BETWEEN 22.5 AND 67.5
\end{minted}

\begin{figure}
    \centering
    \includegraphics[width=.5\linewidth]{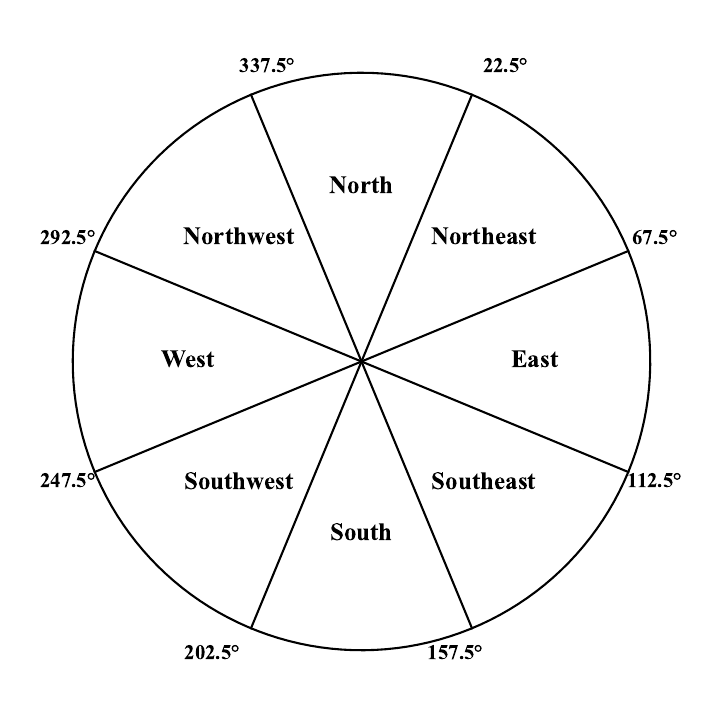}
    \caption{Direction angle ranges}
    \label{fig:direction_circle}
\end{figure}

\noindent\textbf{\textit{Towards:}} For this predicate, an additional anchor point is added that determines the direction in which the user is interested. The angle is computed with the following SQL:
\begin{minted}
[
fontsize=\footnotesize
]{postgresql}
degrees(ST_Azimuth(ancrhor_point1, anchor_point2))
\end{minted}
The geometries are then filtered similar to the direction predicate, but the angle range is based on the computed angle $\pm 22.5$.

\noindent\textbf{\textit{Intersects:}} This type of predicate takes a defined region as its anchor, like the boundaries of a city or a state, instead of anchoring points as the previous predicates. The intersection predicate selects the objects that intersect with the provided boundary, and we use it in questions that ask for aggregate values.
\blue{We note that the English phrasing ``in [region]'' (as used in templates T27 and T28) may more naturally suggest a containment semantic. However, we use \texttt{ST\_Intersects} rather than \texttt{ST\_Contains} because some spatial objects (e.g., rivers, long roads) may cross region boundaries, and we chose to include objects that partially overlap with the region. For example, there are roads that can cross state boundaries. Future versions of the benchmark could include separate templates for Contains vs.\ Intersects to test whether QA systems can distinguish between these operations.} \rev{R1.5}

The \textit{direction} and \textit{towards} predicates are combined with one of the \textit{nearest neighbor} or \textit{range} predicates. As such, in total, we have seven different spatial selectors. Although there are many predicates that we can consider, we selected these five predicates to generate a diverse set of spatial questions. More variety can be easily added in the future, for example, to consider more topological operations such as contains, overlaps, touches, etc.

\bigskip
\noindent\underline{Non-Spatial Predicates}

\noindent These provide an additional aspect to create more complex questions. They are based on other attributes that exist in our tables that we defined in Section~\ref{sec:ref_data}. We only selected non-spatial predicates for points representing restaurants, museums, and hospitals. For museums, the predicate is to specify the type of museum in which the user is interested. For restaurants, the predicates are about the cuisine, but also about services such as delivery, drive-through, or outdoor seating. And for hospitals, we added a predicate for filtering for hospitals that provide emergency services. We can easily add many more predicates of this type, but the ones that we selected provide us with enough variety to generate many natural-sounding questions. This component is used to create four additional templates from T1, T5, T13, and T17. These are templates T2, T6, T14, and T18. The difference is in how POI\_CAT is instantiated. For example, with the inclusion of a non-spatial predicate that specifies the type of museum, it would be instantiated with the phrase \mintinline[fontsize=\footnotesize,breaklines]{python}{'art museum'}, and the SQL will include this predicate \mintinline[fontsize=\footnotesize,breaklines]{SQL}{museum = 'art'}.

\bigskip
\noindent\underline{Output Types}

\noindent The output type or the answer type also adds another aspect to the questions. The user may ask about the name of an entity based on the provided predicates or about the location of the desired entity, such as the coordinates or the address. Additionally, the outputs can also be obtained from spatial operators such as distance, area, and length, which can also be aggregated according to the question.

\noindent\textbf{\textit{Entity name:}} The display name of the spatial object that matches the question.

\noindent\textbf{\textit{Location:}} The coordinates or the complete address of the spatial object that matches the question.

\noindent\textbf{\textit{Direction:}} The azimuth angle in degrees toward which the spatial object that matches the predicates of the question can be found starting from the anchor point of the question and where zero represents the north direction.

\noindent\textbf{\textit{Distance:}} The distance at which the spatial object that corresponds to the predicates of the question can be found, starting from the anchor point of the question.

\noindent\textbf{\textit{Count:}} The number of spatial objects that match the question according to the topological operation it requires.

\noindent\textbf{\textit{Area:}} The area in meters squared of the spatial object or objects that match the predicates of the question.

\noindent\textbf{\textit{Length:}} the length in meters of the spatial object or objects that match the predicates of the question.

\noindent\textbf{\textit{External:}} external responses involve information retrieved from outside the system (out-of-schema) after determining the spatial objects that answer the question. Examples include the capacity, the year it was created, the person who designed, etc.

\bigskip
\noindent\noindent\underline{Aggregate operations}

\noindent Some templates require an aggregation step before producing the final answer. These templates include those that ask for the total area or length, or the count, as well as the templates that ask about the entity that has the maximum area or length.

\subsection{Open Retrieval}\label{sec:out_of_schema}

We also create a set of multi-source questions, which involve information that is not part of our reference data described in \autoref{sec:ref_data}. This adds another layer of complexity to these questions.

We create this by selecting one of the entities in a generated question, which could be an anchor POI or the entity that represents the answer. Then, we retrieve some information from Wikipedia about it, such as the year it was built, who built it, among several other attributes.

We create two types of multi-source questions. The first type involves asking about out-of-schema information for the entity of the answer, like T7 in \autoref{tab:question_templates}. There are several ways to phrase the question in this case, for example, \mintinline[fontsize=\footnotesize]{python}{'What is the nearest hospital from ...?'}
can be changed to \mintinline[fontsize=\footnotesize]{python}{'What type of emergency department is available at ...?'}. So, to answer such a question, first the nearest hospital must be determined, then out-of-schema information about it must be obtained.

The second type of multi-source question involves modifying the name of the anchor point with out-of-schema information; see T8 in \autoref{tab:question_templates}. One such example is that when the anchor point is a university, then its name can be replaced by \mintinline[fontsize=\footnotesize]{python}{'the university with ... as its mascot'}.

These two types add more complexity, since the answer cannot be directly obtained from the reference data. And require multiple steps and synthesizing information from multiple sources to obtain the correct answer.

Although we only included these two types, it is possible to add more variety based on them. For example, region names can be replaced with descriptions about something that uniquely identifies them. Also, instead of just using Wikipedia, other sources can be used. However, we believe that the two types that we included and the attributes we considered are sufficient to test the ability of QA systems to answer questions that involve multiple steps, especially since selecting suitable attributes and ensuring that they translate to natural-sounding questions involves some manual work. Also, later we show that even at this level of complexity the baselines we consider fail in answering these types of questions.

Next, we discuss the methodology for instantiating a question from a template and identifying the correct answers.

\subsection{Question Generation}\label{sec:questionsgen}

\begin{figure}
    \centering
    \includegraphics[width=.3\linewidth]{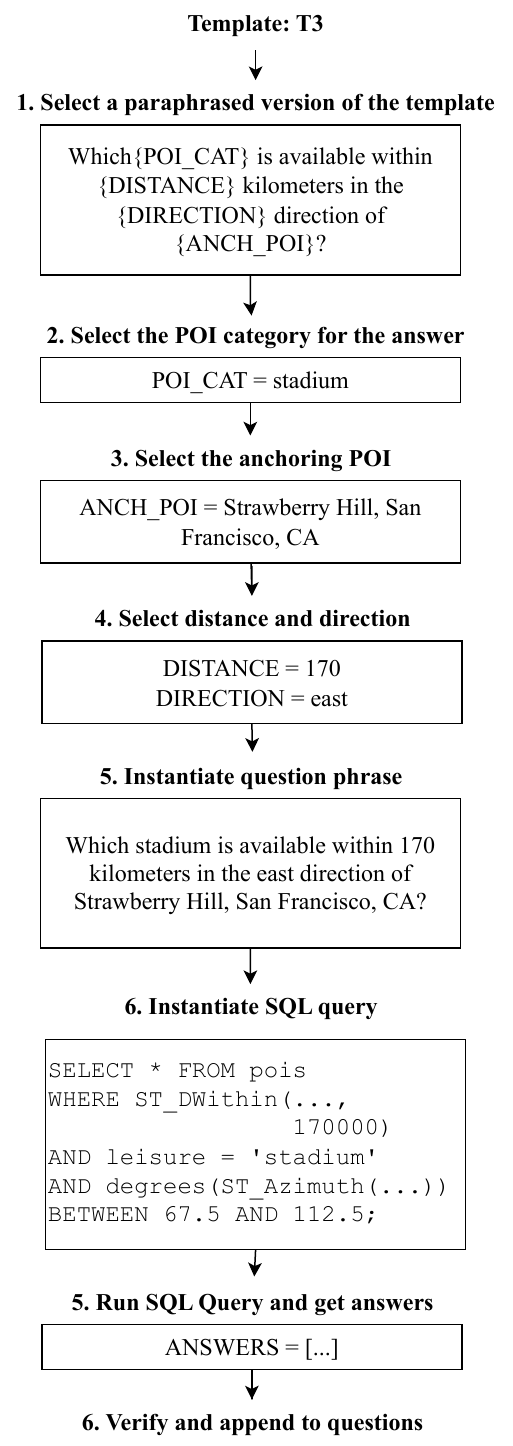}
    \caption{Example of Generating a Question from a Template}
    \label{fig:question_generation}
\end{figure}

In this section, we discuss how to generate questions from a template. The complete process is shown in \autoref{fig:question_generation}, which shows an example of a question instantiated using the T3 template from \autoref{tab:question_templates}. Each template has a fixed set of possible phrases in addition to the one shown in \autoref{tab:question_templates}. The first step is to select a text phrase randomly from all the phrases available for the template that were created and stored in a previous step. In this step, we used an LLM to create different phrases for each template of questions. We also manually edit the LLM output to ensure that all text phrases for a template are valid. 
Each template also has an associated SQL template. Once we have the question phrase and the SQL template, we can start with instantiating them.

In the next step, we start by selecting a value for \mintinline{text}{POI_CAT} that the question asks for. The category is selected randomly from a defined list of possible categories. \blue{These parameter lists were curated manually: the \mintinline{text}{ANCH_POI} categories include prominent, well-known POI types (aquariums, attractions, universities, etc.) as listed in \autoref{tab:spatial-objects}; \mintinline{text}{POI_CAT} values are drawn from the same table; \mintinline{text}{DISTANCE} values are sampled uniformly at random from the range $[1, 200]$~km; and \mintinline{text}{DIRECTION} values are sampled uniformly from the eight compass directions. These lists were designed to produce natural-sounding questions about entities that everyday users would typically search for.} \rev{R1.7}
Additionally, for templates with a non-spatial filter, the filter is appended to the category name and added to the SQL query. Once a category and its filter are determined, a POI that matches it is selected randomly. This POI is used to limit our search in the next step. \blue{Specifically, its geographic coordinates define a local region; the anchor POI (\mintinline{text}{ANCH_POI}) is then selected from the predefined category list within this vicinity, a square of $50km$ in each direction from this point. This helps selecting anchroing points that will have a valid answer. This avoids generating questions with empty answer sets, which would significantly slow the question generation process. Note this is only applied to the templates with two anchoring points. Since those can result in many pair of anchoring points that result in an empty answers, without this limitation.} \rev{R2.18}
We do this to ensure that anchor points are chosen in locations with existing answers, as we exclude questions without answers. Without limiting the search space, the generator takes a long time to find questions with answers. We excluded questions without answers; however, evaluating a QA system on such questions could be insightful. However, generating such questions is simple, as we can create them using random names and coordinates. Next, we randomly select a POI for \mintinline{text}{ANCH_POI} from a predefined list of categories and within the restricted region identified above. After that, we randomly select a value for \mintinline{text}{DIRECTION} and \mintinline{text}{DISTANCE}. All parameters are now defined. Next, both the question phrase and the SQL can be instantiated. The question text is also passed to a grammar checker and correction tool because some templates require pluralizing a name or appending the \textit{`a`} or \textit{`an`} articles.

The next step is to run the instantiated SQL query to get the answers. If the SQL yields no answers, the question is skipped, and the process restarts with new values. Otherwise, a verification and quality check step is applied before the question is appended with all its associated entities to a question bank.

For  multi-source questions, the question is modified after identifying relevant out-of-schema information based on its type. This process was described in \autoref{sec:out_of_schema}.

Other templates are instantiated following a similar procedure based on the types of objects in the template.


For each template, in the question generation step, we generate $1000$ questions, with the exception of T7 and T8. A total of $26,000$, all of which have passed this quality check. Additionally, a number of $150$ questions for each multi-source type, T7 and T8, are included, since they are more expensive to generate. In the final benchmark, we include only $100$ questions of each type, which adds up to $2800$ questions. Next, we discuss the process of how these questions were selected from the larger set.

Using a question template that has \mintinline{text}{POI_CAT} and \mintinline{text}{ANCH_POI} as parameters, such as T5, we can divide its $1000$ questions into hundreds of smaller sets. For example, in this case all the questions that ask about a ``restaurant'' (\mintinline{text}{POI_CAT}), and starting from a ``park'' (\mintinline{text}{ANCH_POI.category}), will be grouped together. Given the variety in all of these parameters, we will have more than $100$ groups, and we select one question from each group that enhances the variety and richness of the questions included in the benchmark. We were able to perform this selection by identifying a set of parameters that allow the questions to be grouped into more than $100$ groups for each template. 

\subsection{Quality Checks}\label{sec:quality}

This section outlines how we check the quality of benchmark questions. In performing quality checks, we make an attempt to ensure that the questions are reasonable. That they can be asked by average users and are not ambiguous to a human reader. For this reason, we only consider more prominent records when instantiating a question. One reason for requiring this is that it helps us avoid inaccurate or incomplete records, since OSM is crowd-sourced.
This verification step involves multiple types of checks. For POIs, we require that they have a name and some address attributes. While it is possible to geoencode a point's coordinates to an address, we use this as a simple indicator of prominent POIs, since a large percentage of POIs in OSM do not have these fields set. For other types of objects, like regions, we require that they have a Wikipedia page available if they are used in anchoring a question, since a lot of regions are administrative regions, and may not even have a name. A region with a Wikipedia page ensures that we select more prominent regions. Additionally, for questions that have a number as their answer such as length, distance, area, and count, we ensure that the answer is not zero.



Finally, we manually review about $10\%$ of the questions for each template with their answers to ensure their correctness.

\subsection{Keeping GS-QA Up-to-date}\label{sec:update}

 It is important to keep the benchmark based on the most recent reference data, because geospatial data is constantly being updated. Since we are using OSM as our main source, in this section we discuss how the benchmark can be updated to match the most recent version of OSM.

For every question that we included in the benchmark, we store with it the OSM records that were used in building the question. Each OSM record has a unique identifier. The process is straightforward and starts by obtaining the most recent version of OSM, and rebuilding our database using it. Next, for every question in the benchmark, we check if any of its associated records changed and update the values if needed. The next step is rerunning the SQL query again to get the answers and update them if needed. Then, the question must also pass the quality checks that we described earlier. For questions that do not pass the quality check, we can discard them and include a new generated question from the same template and possibly based on the same categories but associated with new OSM records.


\section{Baselines}
\label{sec:baselines}

To test the validity of our benchmark, \blue{we built three types of baselines} to evaluate how they perform on it. The purpose of building these baselines is three-fold. First, they allow us to understand how the state-of-the-art LLMs behave with the benchmark, which can reveal interesting research problems. Second, the results of these baselines will allow future, more efficient techniques to be compared with these results. Third, they provide an example of how the benchmark can be used and verify its function.

We define {\em three types} of baselines and each is tested with \blue{{\em three LLMs}, two proprietary and one open-source, for a total of {\em nine baselines}}. We also add an additional baseline that is based on a random answer generator. The first type of baselines is based on feeding a question directly to an LLM without any context. The second type of baseline incorporates a Text2SQL step that takes the question and possibly some context about the tables in the database to translate the question into an SQL query first. Then, the SQL query is passed to our database, and the answer is used as additional context for the model to provide the final answer to the question in a textual form. The third type involves Retrieval Augmented Generation (RAG). In this method, a datastore is used to maintain reference data. Before passing a question to the LLM, a retrieval step is performed to get the most relevant records for the question from the datastore. The retrieved records are appended to add context to the LLM to help enhance the final answer. In total, we have \blue{nine} baselines summarized in \autoref{tab:baselines}, and each is discussed in detail in the following subsections.

\begin{table}[]

\caption{\blue{Baseline combinations}}
\label{tab:baselines}
\footnotesize
\begin{tabular}{|l|l|l|l|}
\hline
\textbf{Label} & \textbf{LLM} & \textbf{\underline{T}ext2SQL}         & \textbf{\underline{R}etrieval}        \\ \hline
M              & \underline{M}inistral-3       & \texttimes & \texttimes \\ \hline
G              & \underline{G}PT4-o       & \texttimes & \texttimes \\ \hline
S              & \underline{S}onnet-4.6    & \texttimes & \texttimes \\ \hline
MT              & \underline{M}inistral-3       & \checkmark & \texttimes \\ \hline
GT             & \underline{G}PT4-o       & \checkmark & \texttimes \\ \hline
ST             & \underline{S}onnet-4.6    & \checkmark & \texttimes \\ \hline
MR              & \underline{M}inistral-3       & \texttimes & \checkmark \\ \hline
GR             & \underline{G}PT4-o       & \texttimes & \checkmark \\ \hline
SR             & \underline{S}onnet-4.6    & \texttimes & \checkmark \\ \hline

R             & \underline{R}andom    & \texttimes & \texttimes \\ \hline
\end{tabular}%
\end{table}

\subsection{Bare LLM Baselines}

\blue{We selected three popular models for all our baselines, two proprietary and one open-source.} These models will help evaluate how state-of-the-art LLMs perform when asked questions from our benchmark, given the various scenarios \blue{in which} different types of context are provided. 
Next, we briefly describe the models that we use.

\blue{\textbf{\textit{Ministral-3:}} Ministral 3 is a family of compact, edge-optimized dense language models released by Mistral AI in December 2025 under the Apache 2.0 license \cite{mistral2025ministral3}. Designed for local and on-device deployment. It is known for its strong cost-to-performance ratio. We use the 14B parameter variant \cite{ollama2025ministral3}.}

\textbf{\textit{GPT-4o:}} It is a transformer model developed by OpenAI and achieved state-of-the-art results in several domains at the time of its release. It can take as input multi-modal data including text, images, and audio, and can generate all three types~\cite{noauthor_gpt-4o_2024}. This model's context window is 128 thousand tokens, and its maximum output size is a little over 16 thousand tokens.

\blue{\textbf{\textit{Sonnet-4.6:}} Claude Sonnet 4.6 is hybrid reasoning model released by Anthropic \cite{anthropic2026sonnet46}. It is well-suited for long-context document analysis, complex multi-step reasoning, and agentic workflows.}



In the first \blue{three baselines, M, G and S}, as shown in \autoref{tab:baselines}, we use prompt engineering to feed the question to the LLM without any additional work. The prompts are designed to provide a concise answer in a standard format that is expected by the benchmark and \blue{that is} easy to evaluate. The following is the system prompt:
\begin{minted}[fontsize=\small,breaklines=true]{text}
Answer the provided user question while satisfying the following requirements:
1. do not include any parts of the question in the answer you must provide the answer directly.
2. provide only the property the user is asking for, like name of an entity, its location, distance, direction.
3. don't provide information the user didn't ask for.
4. any number must be written as words and rounded to the nearest ten.
5. only use metric units.
\end{minted}
We found that this prompt provides concise answers that are relevant to the questions, which are more suitable for text-based evaluation discussed in Section~\ref{sec:eval}.


\subsection{Text2SQL Baselines}

\begin{figure}
    \centering
    \includegraphics[width=.3\linewidth]{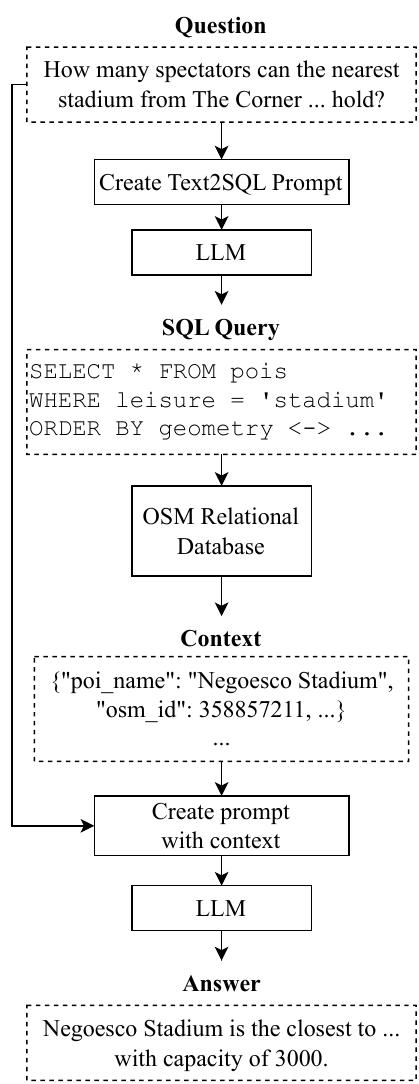}
    \caption{Question Answering Pipeline with Text2SQL}
    \label{fig:baselines_text2sql}
\end{figure}

This baseline uses an LLM to answer questions in three steps. First, it uses the LLM as a Text2SQL generator to create a SQL query out of the question. This step also feeds the database schema to guide the generation process. Second, it runs the produced SQL query on a traditional database to get the answer. Finally, it feeds the SQL result as context to the LLM to produce the final answer in the desired format.

The Text2SQL problem using LLMs has recently received a lot of attention. There are various benchmarks specifically for this task, like~\cite{text2sql_vldb_2024,li_can_2023,zhongSeq2SQL2017}. Although our objective is not to provide a Text2SQL benchmark, our benchmark can be specifically used to evaluate queries that include spatial components. We will include this evaluation later in the experiments. \autoref{fig:baselines_text2sql} shows the complete QA pipeline when Text2SQL is used. The Text2SQL prompt includes the user's question, and context that includes information about the database schema like the table names, and information about each table and its columns. \blue{For example, the schema context includes entries such as: ``Table: \texttt{pois}; Columns: \texttt{geometry} (Point), \texttt{name} (text), \texttt{amenity} (text, values include `restaurant', `hospital'), \texttt{tourism} (text, values include `hotel', `museum'), \texttt{leisure} (text, values include `park', `stadium').'' This provides the model with better context about each table, and the values they contain.}  \rev{R1.8}
After passing the prompt to the LLM, we extract the SQL query produced by the LLM, and execute it on our database. If the query is valid and produces an output. This output is used as context to get the final answer to the user's question. The baselines \blue{MT, GT and ST} are based on this pipeline.

\blue{We note that the Text2SQL baselines do not have access to Wikipedia data, which means they cannot be expected to correctly answer multi-source questions (T7, T8) that require external information. A more complete baseline would combine Text2SQL with a Wikipedia retrieval component, potentially using tool-call mechanisms. We leave this as an important direction for future baseline design.} \rev{R2.17}

\subsection{Retrieval Augmented Generation Baselines}

\begin{figure}
    \centering
    \includegraphics[width=.3\linewidth]{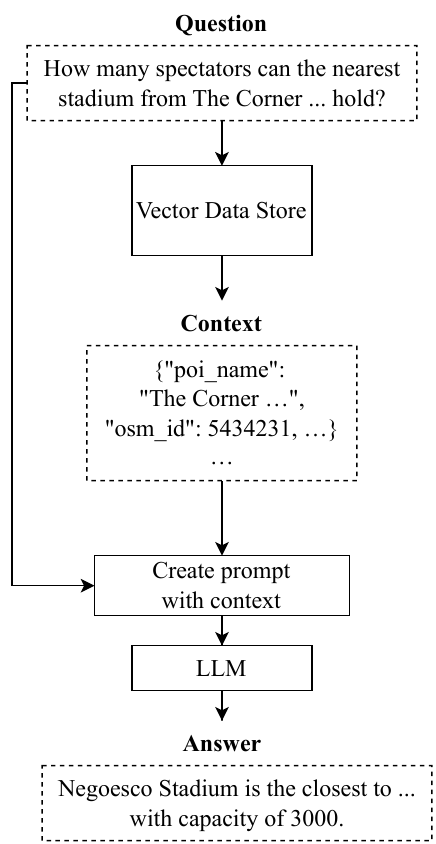}
    \caption{Question Answering Pipeline with RAG}
    \label{fig:baselines_rag}
\end{figure}

One way to improve the quality of the generated answers is to use {\em Retrieval Augmented Generation} (RAG) \cite{10.5555/3495724.3496517}. This method works by first running a similarity search to retrieve the most similar documents to the question and then uses them as context to answer the question by the LLM. Note that technically the previous Text2SQL pipeline can also be considered a RAG method, but here we mainly refer to dense-retrieval methods \cite{dense_retreival_survey}. In dense-retrieval, the search is based on embeddings that represent some text. In our case, we use the \blue{Nomic Embed \text{nussbaum2024nomic} model provided in \cite{ollama2024nomicembed} to generate the text embeddings. It is an open-source model that provides reproducible embeddings.} \blue{Each database record is embedded as a text string containing the entity's name, type/category, and address attributes. Geometry coordinates are not directly included in the text representation.} \rev{R2.21}
A datastore is needed for an efficient RAG-based baseline. We build it using the \blue{Nomic} model embeddings, and use ChromaDB~\cite{noauthor_chroma-core/chroma_2025}, an open-source database for storing those embeddings, and performing efficient searches. We create our datastore by including all records in all the tables in the reference database that we used when generating the questions, and a small subset of Wikipedia, containing those pages that were considered when generating the  multi-source questions.


To answer a question, we first encode it using the same \blue{Nomic embedding} model, and then we use ChromaDB to get the top-10 records with the most similar embeddings. After that, we prompt the LLM with the question and provide the closest documents as context to get the final answer. The baselines \blue{MR, GR and SR} are based on this pipeline.

\section{Experiments}
\label{sec:experiments}


In this section, we study the effectiveness of the baselines described in Section~\ref{sec:baselines} on the GS-QA benchmark.
Simply applying standard text-based evaluation techniques, comparing the answer string to the ground truth string, is not adequate, given the inherent structure of spatial answers. For example, two answers may not share any words, but they may be spatially too close to each other. For that, we consider a comprehensive suite of spatial-aware evaluation measures, in addition to standard text-based measures, as discussed in Section~\ref{sec:eval}. The results are presented in Section~\ref{sec:results} and a discussion in Section~\ref{sec:discussion}.

\subsection{Evaluation Strategy}
\label{sec:eval}

We use two evaluation approaches: \textit{text-based matching}, and \textit{structured spatial-aware evaluation}.

The text-based matching searches for common words between the ground truth and the generated answers. Before matching, a simple preprocessing step is applied to remove punctuation, convert to lower case, and convert numbers to words. Then, three values are computed. Precision is the percentage of common words between the two sentences divided by the number of words in the predicted sentence.
Recall is the percentage of common words divided by the number of words in the ground truth answer.
The F1 score is the harmonic mean of the two scores. When precision and recall are both zero, we also set F1 to zero.
\ms{We do not consider scores like BERTScore~\cite{bert-score}, since when evaluating addresses or entity names exact matching is required as compared to semantic similarity.}

For the structured spatial-aware evaluation, the answer is first converted to a JSON document, using an \blue{LLM known to work well with programming tasks} and a prompt specifying the desired JSON schema, as in the following example: 
\begin{minted}[fontsize=\small]{text}
The nearest science museum is the Orlando Science Center,
located in Orlando, Florida, approximately fifty kilometers
north of the Central Florida Zoo & Botanical Gardens,
Sanford, FL.
\end{minted}
Converted to JSON:
\begin{minted}[fontsize=\small]{json}
{
  "name": "Orlando Science Center",
  "address": "Orlando, Florida",
  "distance": 50, "azimuth_angle": 90,
  "count": null, "length": null, "area": null
}
\end{minted}

Note that inaccuracies might occur in this step, such as setting the angle to $90^\circ$ for the north direction when north is $0^\circ$. \blue{To minimize this risk, we tested a few LLM options for the parsing step and found Qwen 3.5 \cite{ollama2025qwen35} to work well. It is optimized for programming tasks and is very efficient and works locally.} Also, \blue{the prompt can contain} additional attributes such as "architect" \blue{which} might be included for  \blue{multi-source} questions.
Based on the JSON fields, we consider spatial-aware measures, customized for each output type, as shown in \autoref{tab:evaluation-metrics}.
Specifically, when the answer type is an entity name, we use the same metrics as in the free-text form. 
We expect that the scores in this case will be better because of the more compact representation compared to the text-based matching approach.
Similarly, \blue{multi-source} answers are evaluated as text using the same metrics.

When the answer type is a location, it is evaluated in two ways. First, the address text is the three text-based measures. In addition, we calculate the distance in meters between the geo-encoded position of the predicted address and the expected position. The distance is divided by a threshold, which we set to five hundred kilometers. If the distance is more than the threshold, the error is set to the maximum value of one. Also, given this threshold, an error of $0.01$ is equivalent to five kilometers in distance. We use the Open Street Map free geoencoding service, called Nominatim~\cite{noauthor_nominatim_nodate}. This step may not always be accurate.




When the answer type is an angle, we evaluate using both an exact match text approach and a geometric metric. For the text-based approach, the angles are converted to their description, and if both angles fall in the same range, they will be an exact match. Angle  ranges are obtained by dividing the $360^\circ$, into eight equal ranges, as shown in Figure~\ref{fig:direction_circle}. For the angle value, an angle error score is computed using the following equation, where zero is an exact match and one is a $180$ degrees difference.

\begin{equation}
    \text{Angle Error} = 
    \begin{cases} 
    \frac{|\text{angle1} - \text{angle2}|}{180}, & \text{if } |\text{angle1} - \text{angle2}| < 180 \\ 
    \frac{360 - |\text{angle1} - \text{angle2}|}{180}, & \text{otherwise} 
    \end{cases}
    \label{eq:angle_error}
\end{equation}

For all other output types, which are numeric, the relative error is used to measure the quality using the following equation.

\begin{equation}
    \text{Relative Error} = \frac{|\text{Prediction}-\text{Actual}|}{Actual} \label{eq:common_words_ratio}
\end{equation}

For all questions in the benchmark, the actual answer for the output types Area, Length, Distance, and Count is never zero, which makes the relative error defined for all questions. For all output types, when the attribute related to it is missing or is out of the expected range, the question is counted as not attempted, such as when the address value is missing or empty and the output type is location. Additionally, the geoencoding step may not produce an output, which might indicate that the address is malformed, and these questions are also counted as not attempted for the distance error score.

For questions that can have multiple valid answers, like questions with the \textit{Range} predicate, each predicted answer is compared to all possible answers, and only the best scores are reported. 

\ms{There are some considerations for questions that can have multiple valid answers. One consideration is to rank all the possible answers based on some quality metric. For example, when there are many restaurants that fit the criteria of the questions, we give more weight to more popular ones. Furthermore, some types of questions can be interpreted differently. For example, users asking about lakes may only be interested in natural lakes, while the system may consider man-made lakes, and seasonal lakes. We leave evaluation of these special cases for future work.}

\begin{table}[]
\caption{Metrics used for parsed output}
\footnotesize
\label{tab:evaluation-metrics}
\begin{tabular}{|l|l|l|l|l|l|}
\hline
\textbf{Output Type}  & \textbf{{\begin{tabular}[c]{@{}l@{}}Recall\end{tabular}}}                & \textbf{{\begin{tabular}[c]{@{}l@{}}F1 \end{tabular}}}                                       & \textbf{\begin{tabular}[c]{@{}l@{}}Distance\\ Error\end{tabular}} & \textbf{\begin{tabular}[c]{@{}l@{}}Angle\\ Error\end{tabular}} & \textbf{\begin{tabular}[c]{@{}l@{}}Relative\\ Error\end{tabular}} \\ \hline
Entity name &  \checkmark &  \checkmark &                                                                   &                                                                &                                                                   \\ \hline
Location &  &  \checkmark &  \checkmark                 &                                                                &                                                                   \\ \hline
Direction  &  &  \checkmark &                                                                   &  \checkmark              &                                                                   \\ \hline
Area                                                              &                           &                                                   &                                                                   &                                                                &  \checkmark                 \\ \hline
Length                                                            &                           &                                                   &                                                                   &                                                                &  \checkmark                 \\ \hline
Distance                                                         &                           &                                                   &                                                                   &                                                                &  \checkmark                 \\ \hline
Count                                                           &                           &                                                   &                                                                   &                                                                &  \checkmark                 \\ \hline
\end{tabular}
\end{table}

\subsection{Evaluation Results}
\label{sec:results}

In the experiments, we organize the results by output types, as shown in Table~\ref{tab:question_templates}, because each type may have its own unique evaluation measures (Table~\ref{tab:evaluation-metrics}).
\blue{The different output types require fundamentally different evaluation approaches because they are not comparable. Evaluating an entity name requires different considerations as compared to evaluating an address, and similarly for other types like direction, distance, and others. We use the most appropriate metric for each output type. As a general metric, we use the precision, recall, and F1 for the free text answers, and compare all the baselines based on the average of these values for all questions in the benchmark. We avoid using LLMs for evaluation since this still requires calibration of the quality of the evaluation, which can only be done through existing evaluation metrics or by human evaluators given a consistent evaluation criteria. Challenges related to using LLMs for evaluation have been discussed in \cite{zheng2023judging, gu2024survey, li2024generation}. We leave more sophisticated evaluation metrics for each type including human evaluation, and LLM evaluation for future work.} \rev{R1.9, R2.20}

{\bf Attempted questions} 
We mark a question as attempted if the generated JSON contains a value for the required key, e.g., for the key ``address,'' ``name'' or ``distance.''
We provide a summary of the percentage of questions attempted   in Table~\ref{tab:attempted-percent}. We see that numeric questions are less likely to be  answered, which we discuss in more detail for each output type. 
\blue{We also notice that the baselines with more context provided namely Text2SQL and RAG, attempt less questions. The model typically responds that it doesn't have the relevant information. We see this as a positive sign as compared to providing a random answer.}



\begin{table}[]
\centering
\caption{\blue{Percentage of attempted questions in parsed answers}}
\label{tab:attempted-percent}
\footnotesize
\begin{tabular}{lrrrrrrrrrr}
\hline
Output type & M & MT & MR & S & ST & SR & G & GT & GR & R \\
\hline
Name & 0.98 & 0.90 & 0.91 & 0.87 & 0.67 & 0.67 & 0.97 & 0.92 & 0.88 & 0.92 \\
Loc & 0.92 & 0.85 & 0.81 & 0.80 & 0.75 & 0.47 & 0.92 & 0.89 & 0.82 & 0.99 \\
Direction & 0.99 & 0.99 & 0.94 & 0.92 & 0.82 & 0.30 & 1.00 & 0.99 & 0.87 & 1.00 \\
Area & 1.00 & 0.81 & 0.65 & 0.92 & 0.33 & 0.28 & 1.00 & 0.95 & 0.84 & 1.00 \\
Count & 1.00 & 0.97 & 0.96 & 0.68 & 0.99 & 0.70 & 1.00 & 1.00 & 0.99 & 1.00 \\
Distance & 1.00 & 0.95 & 0.85 & 0.93 & 0.96 & 0.80 & 0.99 & 0.98 & 0.96 & 1.00 \\
Length & 0.85 & 0.71 & 0.62 & 0.83 & 0.43 & 0.16 & 0.84 & 0.84 & 0.81 & 0.88 \\
\hline
\textbf{Avg} & 0.96 & 0.88 & 0.82 & 0.85 & 0.71 & 0.48 & 0.96 & 0.94 & 0.88 & 0.97 \\
\hline
\end{tabular}
\end{table}

{\bf \blue{Free text output evaluation}} 
\blue{We use the F1 score on the free text responses to evaluate all the baselines across a single score. We notice that the baseline ST performs the best on average. This also correlates with the results using the specific metrics for each type. Also, notice here that this evaluation considers all free text answers including the answers that we mark unattempted. This can explain the low F1 score for ST on the area output type, since it only attempted $28\%$ of questions. The average score also correlates well with other values showing that the Text2SQL baselines generally perform better than other baselines.} \rev{R1.9}

\begin{table*}
\caption{\blue{F1 score for free text output}}
\label{tab:prf_text_average}
\footnotesize

\begin{tabular}{l|cccccccccc}
\hline
Type & M & MT & MR & S & ST & SR & G & GT & GR & R \\
\hline
Name & 0.09 & 0.14 & 0.10 & 0.10 & 0.33 & 0.09 & 0.13 & 0.29 & 0.11 & 0.11 \\
Loc & 0.02 & 0.02 & 0.01 & 0.06 & 0.14 & 0.02 & 0.03 & 0.05 & 0.01 & 0.05 \\
Angle & 0.07 & 0.12 & 0.07 & 0.06 & 0.34 & 0.01 & 0.11 & 0.12 & 0.11 & 0.07 \\
Area & 0.17 & 0.17 & 0.09 & 0.10 & 0.09 & 0.05 & 0.14 & 0.18 & 0.14 & 0.42 \\
Count & 0.10 & 0.20 & 0.06 & 0.03 & 0.32 & 0.02 & 0.08 & 0.20 & 0.02 & 0.11 \\
Distance & 0.09 & 0.16 & 0.13 & 0.04 & 0.24 & 0.12 & 0.06 & 0.13 & 0.12 & 0.23 \\
Length & 0.17 & 0.19 & 0.11 & 0.12 & 0.13 & 0.02 & 0.15 & 0.19 & 0.14 & 0.33 \\
\hline
\textbf{Avg} & 0.10 & 0.14 & 0.08 & 0.07 & 0.23 & 0.05 & 0.10 & 0.17 & 0.09 & 0.19 \\
\hline
\end{tabular}

\end{table*}

{\bf Return type: Entity name}
We show the results for twelve templates from  Table~\ref{tab:question_templates} in Table~\ref{tab:entity-name-eval}. 
We consider two scores. For the free text output, we look at the recall ($R$) score. For the parsed output, we use the F1 value to compare the true answer to the extracted entity name or the  multi-source attribute in the case of T7. \ms{This score only includes the attempted questions as defined earlier.} 
We notice that the baseline \blue{ST} has considerably higher scores for several question categories.  


\ms{Moreover, we manually checked many of the answers, and noticed a lot of the answers are correct, and it correlates with the scores shown in the table.} Also, one observation is that the scores are affected by common words in the entity names. For example, many parks and lakes have the words 'Park' and 'Lake' in their official names. Therefore, if the official name has only two words, the presence of these words would influence precision and recall at $50\%$. On the other hand, some names are spelled a little differently than the reference answer, which results in the penalization of correct answers. With regard to multi-source questions, T7 and T8, we noticed that \blue{almost all questions were answered incorrectly or are unattempted}.

\begin{table}[]
\centering
\caption{\blue{Evaluation of templates with entity name}}
\label{tab:entity-name-eval}
\footnotesize
\setlength{\tabcolsep}{2pt}

\begin{tabular}{l|cccccccccc|cccccccccc}
\hline
\multirow{2}{*}{TID} & \multicolumn{10}{|c|}{Recall on full text output} & \multicolumn{10}{c}{F1 on parsed output} \\
\cline{2-21}
\multicolumn{1}{l|}{} & M & S & G & MT & ST & GT & MR & SR & GR & R & M & S & G & MT & ST & GT & MR & SR & GR & R \\
\hline
T1 & 0.04 & 0.06 & 0.04 & 0.05 & 0.33 & 0.11 & 0.02 & 0.04 & 0.03 & 0.02 & 0.37 & 0.49 & 0.39 & 0.46 & 0.81 & 0.74 & 0.32 & 0.33 & 0.33 & 0.26 \\
T2 & 0.09 & 0.17 & 0.11 & 0.12 & 0.72 & 0.27 & 0.08 & 0.12 & 0.06 & 0.08 & 0.25 & 0.33 & 0.26 & 0.35 & 0.80 & 0.64 & 0.26 & 0.24 & 0.22 & 0.22 \\
T3 & 0.08 & 0.14 & 0.08 & 0.07 & 0.53 & 0.16 & 0.04 & 0.08 & 0.05 & 0.04 & 0.24 & 0.24 & 0.24 & 0.24 & 0.71 & 0.47 & 0.19 & 0.20 & 0.24 & 0.14 \\
T4 & 0.04 & 0.06 & 0.02 & 0.03 & 0.25 & 0.03 & 0.02 & 0.03 & 0.02 & 0.02 & 0.40 & 0.32 & 0.41 & 0.36 & 0.45 & 0.41 & 0.32 & 0.30 & 0.33 & 0.35 \\
T5 & 0.12 & 0.18 & 0.12 & 0.33 & 0.78 & 0.66 & 0.10 & 0.14 & 0.11 & 0.11 & 0.10 & 0.15 & 0.11 & 0.31 & 0.84 & 0.65 & 0.10 & 0.14 & 0.11 & 0.09 \\
T6 & 0.15 & 0.19 & 0.17 & 0.18 & 0.44 & 0.57 & 0.13 & 0.17 & 0.13 & 0.19 & 0.13 & 0.14 & 0.17 & 0.17 & 0.47 & 0.56 & 0.13 & 0.16 & 0.14 & 0.19 \\
T7 & 0.09 & 0.33 & 0.02 & 0.12 & 0.30 & 0.14 & 0.08 & 0.16 & 0.04 & 0.19 & 0.02 & 0.04 & 0.03 & 0.03 & 0.15 & 0.09 & 0.02 & 0.05 & 0.03 & 0.04 \\
T8 & 0.13 & 0.19 & 0.10 & 0.16 & 0.30 & 0.14 & 0.06 & 0.11 & 0.06 & 0.10 & 0.12 & 0.10 & 0.09 & 0.16 & 0.62 & 0.13 & 0.06 & 0.09 & 0.07 & 0.05 \\
T9 & 0.15 & 0.19 & 0.16 & 0.16 & 0.66 & 0.25 & 0.10 & 0.13 & 0.12 & 0.13 & 0.13 & 0.17 & 0.15 & 0.15 & 0.88 & 0.25 & 0.10 & 0.10 & 0.12 & 0.14 \\
T10 & 0.13 & 0.07 & 0.08 & 0.12 & 0.21 & 0.16 & 0.05 & 0.09 & 0.06 & 0.07 & 0.09 & 0.04 & 0.07 & 0.09 & 0.27 & 0.16 & 0.04 & 0.09 & 0.06 & 0.06 \\
T11 & 0.22 & 0.27 & 0.33 & 0.33 & 0.36 & 0.40 & 0.23 & 0.26 & 0.32 & 0.10 & 0.29 & 0.38 & 0.48 & 0.45 & 0.69 & 0.59 & 0.33 & 0.40 & 0.48 & 0.25 \\
T12 & 0.21 & 0.28 & 0.27 & 0.32 & 0.33 & 0.36 & 0.17 & 0.24 & 0.27 & 0.19 & 0.23 & 0.30 & 0.32 & 0.37 & 0.60 & 0.43 & 0.20 & 0.35 & 0.34 & 0.23 \\
\hline
\textbf{Avg} & 0.12 & 0.18 & 0.12 & 0.17 & 0.43 & 0.27 & 0.09 & 0.13 & 0.10 & 0.10 & 0.20 & 0.22 & 0.23 & 0.26 & 0.61 & 0.43 & 0.17 & 0.20 & 0.21 & 0.17 \\
\hline
\end{tabular}

\end{table}

{\bf Return type: Location}
The results are provided in Table~\ref{labeltab:loc-eval}. Looking at the distance error score, we see that several templates have errors noticeably better than the random baseline. We observed that many valid addresses were provided. However, there are a few cases where the baseline did not provide an address, and in the parsing step, the LLM selected the address of the ANCH\_POI from the question instead of the answer. 

\begin{table}[]
\centering
\caption{\blue{Evaluation of templates with location}}
\label{labeltab:loc-eval}
\footnotesize
\setlength{\tabcolsep}{2pt}

\begin{tabular}{l|cccccccccc|cccccccccc}
\hline
\multirow{2}{*}{TID} & \multicolumn{10}{c|}{F1 for address text} & \multicolumn{10}{c}{Distance Error} \\
\cline{2-21}
\multicolumn{1}{l|}{} & M & S & G & MT & ST & GT & MR & SR & GR & R & M & S & G & MT & ST & GT & MR & SR & GR & R \\
\hline
T13 & 0.26 & 0.36 & 0.30 & 0.29 & 0.41 & 0.37 & 0.32 & 0.34 & 0.33 & 0.32 & 0.25 & 0.12 & 0.18 & 0.23 & 0.18 & 0.22 & 0.26 & 0.27 & 0.23 & 0.12 \\
T14 & 0.20 & 0.28 & 0.21 & 0.19 & 0.44 & 0.30 & 0.23 & 0.18 & 0.23 & 0.24 & 0.22 & 0.14 & 0.21 & 0.20 & 0.18 & 0.19 & 0.36 & 0.45 & 0.35 & 0.21 \\
T15 & 0.19 & 0.27 & 0.21 & 0.19 & 0.41 & 0.24 & 0.23 & 0.23 & 0.22 & 0.21 & 0.30 & 0.18 & 0.27 & 0.24 & 0.19 & 0.26 & 0.18 & 0.23 & 0.20 & 0.12 \\
T16 & 0.42 & 0.46 & 0.35 & 0.39 & 0.47 & 0.42 & 0.39 & 0.37 & 0.41 & 0.24 & 0.14 & 0.09 & 0.11 & 0.17 & 0.25 & 0.13 & 0.29 & 0.44 & 0.26 & 0.89 \\
T17 & 0.27 & 0.29 & 0.29 & 0.30 & 0.46 & 0.35 & 0.32 & 0.26 & 0.31 & 0.05 & 0.21 & 0.17 & 0.25 & 0.15 & 0.12 & 0.19 & 0.22 & 0.20 & 0.22 & 0.95 \\
T18 & 0.14 & 0.14 & 0.11 & 0.10 & 0.31 & 0.23 & 0.12 & 0.11 & 0.13 & 0.13 & 0.46 & 0.39 & 0.42 & 0.48 & 0.32 & 0.40 & 0.51 & 0.68 & 0.48 & 0.88 \\
T19 & 0.19 & 0.17 & 0.17 & 0.21 & 0.49 & 0.22 & 0.20 & 0.17 & 0.20 & 0.20 & 0.18 & 0.14 & 0.16 & 0.22 & 0.28 & 0.22 & 0.16 & 0.27 & 0.16 & 0.31 \\
T20 & 0.21 & 0.24 & 0.18 & 0.23 & 0.34 & 0.21 & 0.23 & 0.20 & 0.23 & 0.05 & 0.22 & 0.22 & 0.14 & 0.27 & 0.10 & 0.16 & 0.26 & 0.31 & 0.28 & 0.89 \\
\hline
\textbf{Avg} & 0.24 & 0.28 & 0.23 & 0.24 & 0.41 & 0.29 & 0.26 & 0.23 & 0.26 & 0.18 & 0.25 & 0.18 & 0.22 & 0.25 & 0.20 & 0.22 & 0.28 & 0.36 & 0.27 & 0.55 \\
\hline
\end{tabular}

\end{table}

{\bf Return type: Direction}
We show the results in Table~\ref{tab:direction-eval}. Since the angle has a limited range of possible values, we first compute the expected error for a random number generator baseline. We find the expected angle error to be $0.5$, since on average the angles would be at $90^\circ$ difference in either direction. Based on this, we notice that for T22, which is based on the k-nearest-neighbor predicate, \blue{most} angle errors are around the expected error for a random generator, even though there are quite a few questions where the correct direction was predicted, \blue{with the exception of GT and ST, which performed better, keeping in mind they attempted less questions of this type.} For T21, the error is less because the score keeps the best matching prediction to any of the possible correct answers. The expected error depends on the number of possible correct answers and the number of predictions. For example, if we have only one possible correct answer, and three predictions were provided, the expected error for a random generator is around $0.25$. We obtained these expected errors by running a simple simulator. Note that the error is also affected by the number of unattempted questions as defined earlier. Similar to the location-based questions, we think a more sophisticated evaluator would first evaluate the predicted entity and that it exists in the reference data, which could be different from the reference data used for building the benchmark, and that it matches all the specifiers in the question. Only after that can we evaluate its direction relative to the ANCH\_POI in the question.
\begin{table}[]
\centering
\caption{\blue{Evaluation of templates with direction}}
\label{tab:direction-eval}
\footnotesize
\setlength{\tabcolsep}{2pt}

\begin{tabular}{l|cccccccccc|cccccccccc}
\hline
\multirow{2}{*}{TID} & \multicolumn{10}{c|}{F1 for parsed direction} & \multicolumn{10}{c}{Angle Error} \\
\cline{2-21}
\multicolumn{1}{l|}{} & M & S & G & MT & ST & GT & MR & SR & GR & R & M & S & G & MT & ST & GT & MR & SR & GR & R \\
\hline
T21 & 0.57 & 0.69 & 0.65 & 0.68 & 0.95 & 0.84 & 0.60 & 0.77 & 0.66 & 0.58 & 0.18 & 0.15 & 0.16 & 0.15 & 0.05 & 0.08 & 0.20 & 0.12 & 0.17 & 0.17 \\
T22 & 0.09 & 0.13 & 0.14 & 0.18 & 0.65 & 0.52 & 0.10 & 0.10 & 0.19 & 0.16 & 0.55 & 0.48 & 0.47 & 0.51 & 0.21 & 0.28 & 0.56 & 0.56 & 0.48 & 0.47 \\
\hline
\end{tabular}

\end{table}
{\bf Other return types}
We provide a summary for the remaining templates in Table~\ref{tab:numeric-eval}. Note that we limit the relative error to a maximum of $1.0$. Hence, any baseline with a score of $1.0$ provided answers very far from the correct answer. 

\blue{The baseline ST provided consistently better scores compared to the random baselines. }
\blue{However, } this \blue{can happen} when the provided answer has a small distance, since most of the correct answers have small distances. We think the evaluation of these types of questions should also evaluate the correctness of the entity where the distance is measured. Otherwise, the LLM may provide some random numbers that may seem correct. Note that we store the entire entity in the reference answer, and the associated SQL query to the questions gets the entity as well as the distance, in case the user has their own reference data.

T27 and T28 are analytical queries that require aggregating the area or length of many geometries that intersect with a provided search area. \blue{Also, not that ST which has the best score only attempted a small percentage of questions. As noted earlier, this we consider this to be much better than providing random answers, and this is reflected on the average score of the attempted questions.}

\begin{table}[]
\centering
\caption{\blue{Relative error for templates with numeric answers}}
\label{tab:numeric-eval}
\footnotesize
\setlength{\tabcolsep}{2pt}

\begin{tabular}{l|l|cccccccccc}
\hline
\textbf{Type} & \textbf{TID} & \textbf{M} & \textbf{S} & \textbf{G} & \textbf{MT} & \textbf{ST} & \textbf{GT} & \textbf{MR} & \textbf{SR} & \textbf{GR} & \textbf{R} \\
\hline
\multicolumn{1}{l|}{\multirow{2}{*}{\textbf{Count}}} & \textbf{T23} & 0.84 & 0.72 & 0.71 & 0.58 & \textbf{0.31} & 0.36 & 0.83 & 0.85 & 0.86 & 0.96 \\
\multicolumn{1}{l|}{} & \textbf{T24} & 0.86 & 0.83 & 0.92 & 0.83 & \textbf{0.73} & 0.96 & 0.87 & 0.95 & 0.94 & 0.98 \\
\hline
\multicolumn{1}{l|}{\multirow{2}{*}{\textbf{Distance}}} & \textbf{T25} & 0.94 & 0.97 & 0.93 & 0.76 & \textbf{0.18} & 0.73 & 0.78 & 0.45 & 0.89 & 1.00 \\
\multicolumn{1}{l|}{} & \textbf{T26} & 0.94 & 0.87 & 0.84 & 0.75 & \textbf{0.15} & 0.66 & 0.80 & 0.62 & 0.68 & 1.00 \\
\hline
\multicolumn{1}{l|}{\textbf{Area}} & \textbf{T27} & 0.97 & 0.98 & 0.98 & 0.89 & \textbf{0.51} & 0.78 & 0.94 & 0.68 & 0.84 & 0.96 \\
\hline
\multicolumn{1}{l|}{\textbf{Length}} & \textbf{T28} & 0.93 & 0.93 & 0.90 & 0.68 & \textbf{0.47} & 0.81 & 0.90 & 0.93 & 0.95 & 1.00 \\
\hline
\end{tabular}

\end{table}

\subsection{Discussion}
\label{sec:discussion}
{\bf Analysis of Text2SQL performance}
The Text2SQL pipeline resulted in significant improvements in the answers for \blue{most of the templates}. 
We categorize the generated SQL queries into valid and invalid, and summarize the results in Table~\ref{tab:text2sql-errors}. \blue{Sonnet 4.6 (S), generated valid SQL for almost $90\%$ of the questions, GPT4-o (G) generated valid queries for about $80\%$ of the questions, Minestral 3 (M) generated valid queries for only about $52\%$ of the questions.} Note that a valid query does not necessarily mean the query resulted in retrieving the correct answers. \blue{We estimate the percentage of correct answers based on the output type, for text based answer types, we count it as correct if the F1 score is greater than or equal to $0.5$, and for other metrics its if the value is less than or equal to $0.1$. The best score is $44.2\%$ out of all the benchmark questions.} \rev{R1.11} A small percentage of the queries timed out. \blue{In the execution pipeline, we set the time limit for each query to be $60$ seconds.} Also, we limit the retrieved records to only $20$ records since some of the generated queries result in retrieving a very large set. We further categorize the invalid SQL queries into multiple sub-categories. A lot of errors are caused by improper use of spatial functions and predicates. There are clearly many improvements needed for a more accurate Text2SQL when it comes to geospatial data.

\begin{table}[]
\centering
\caption{Text2SQL error summary}
\label{tab:text2sql-errors}
\footnotesize
\begin{tabular}{l|ccc}
\hline
\textbf{Error Type} & \textbf{M} & \textbf{S} & \textbf{G} \\
\hline
\textbf{Valid SQL} & 51.8\% & 89.7\% & 80.4\% \\
\hdashline
\quad Timeout & 1.6\% & 1.8\% & 1.0\% \\
\quad Correct Output & 20.2\% & 44.2\% & 33.0\% \\
\quad Incorrect Output & 30.0\% & 43.7\% & 46.3\% \\
\hline
\textbf{Invalid SQL} & 48.2\% & 10.3\% & 19.6\% \\
\hdashline
\quad Column not found & 25.9\% & 3.3\% & 5.5\% \\
\quad Sub-query error & 1.7\% & 0.0\% & 0.1\% \\
\quad Function not found & 8.1\% & 3.3\% & 9.9\% \\
\quad Other & 3.3\% & 2.5\% & 3.3\% \\
\quad Missing FROM & 6.2\% & 0.1\% & 0.0\% \\
\quad Syntax error & 1.9\% & 0.0\% & 0.7\% \\
\quad Relation not found & 0.1\% & 0.0\% & 0.0\% \\
\quad Operator not found & 0.9\% & 1.0\% & 0.1\% \\
\hline
\end{tabular}

\end{table}

{\bf Analysis of RAG performance}
From the previous results, \blue{Tables~\ref{tab:prf_text_average}-\ref{tab:numeric-eval}},\rev{R1.12} it is clear that the dense-retrieval-based baselines \blue{MR, SR, and GR} \blue{in many cases perform worst perform worse compared to other baselines}. This is mainly due to the quality of the embeddings, which we found to not give importance to location information. To evaluate this method further, we manually crafted a few examples to test its quality. For example, when we search using non-spatial terms like the type of cuisine, generally relevant records are retrieved. However, when we search using the name of a city usually it does not retrieve relevant records. When combining non-spatial keywords with the name of a region, usually records relevant to the non-spatial keywords are retrieved. Clearly, a more suitable dense-retrieval method that gives importance to geospatial information is needed, since a keyword-based sparse-retrieval might have performed better.


{\bf Summary of results} The results show that only questions with \textit{entity name} output type provided relatively good answers, and only for those questions with the simplest spatial predicate. However, for other answer types, the baselines mostly provide random answers, \blue{with the exception of the baselines with Sonnet 4.6 which is more likely to not attempt answering instead of providing a random answer compared to other models.} We also discussed the existing limitations in the evaluation method. \ms{The baselines we used are based on existing tools and are not tailored specifically for geospatial data, since our objective is proposing the benchmark, and understanding how current LLM based baselines would perform. We think that having these results is encouraging for future research.}

\section{Conclusion and Future Work}
\label{sec:future}

We created an extensible benchmark, GS-QA, for spatial QA, based on an automated question and answer generator. GS-QA includes a variety of spatial predicates, geospatial and non-geospatial entities, and output types. 
We proposed spatial-specific evaluation measures that go beyond standard text-based matching.
We have shown that existing LLM-based baselines are not sufficient to answer such questions.

We have identified several future research directions. 
First, while we have proposed an evaluation strategy that takes spatial characteristics into account, a more sophisticated and standardized evaluation is still needed. For example, we may want to quantify how good the answer ``California" is if the correct answer is ``Los Angeles."
Further, instead of just evaluating based on the final answer, we can also evaluate the correct identification of the relevant geospatial entities.
Moreover, when extracting a spatial entity from an LLM's response, parsing errors may occur, which need to be quantified.
Second, we argue that more advanced GeoQA systems are needed. Such systems must be able to handle a variety of spatial predicates and integrate non-spatial information, like in the case of multi-source questions that we included. 
Third, better retrieval methods that take geospatial information into account are needed, as seen in the discussion for the Text2SQL and RAG results. 
Finally, research is needed to support even more advanced geospatial questions that could be asked by geospatial analysts in areas such as urban planning and epidemiology.



\bibliographystyle{ACM-Reference-Format}
\bibliography{acmart}

\end{document}